\documentclass{article}
\usepackage[round]{natbib}
\usepackage{arxiv}
\usepackage[utf8]{inputenc} % allow utf-8 input
\usepackage[T1]{fontenc}    % use 8-bit T1 fonts
\usepackage{lmodern}        % https://github.com/rstudio/rticles/issues/343
\usepackage[colorlinks=true, linkcolor=blue, citecolor=blue, urlcolor=blue]{hyperref}
\usepackage{url}            % simple URL typesetting
\usepackage{booktabs}       % professional-quality tables
\usepackage{amsfonts}       % blackboard math symbols
\usepackage{nicefrac}       % compact symbols for 1/2, etc.
\usepackage{microtype}      % microtypography
\usepackage{graphicx}
\usepackage{xcolor}
\usepackage{calc}
\usepackage{float}
\usepackage{amsmath}
\usepackage{dsfont}
\graphicspath{ {./Figures/} }

\title{On the importance of tail assumptions in climate extreme event attribution}

\author{
    Mengran Li
   \\
    School of Mathematics and Statistics, University of Glasgow,
Glasgow, the United Kingdom \\
   \\
  \texttt{\href{mailto:m.li.3@research.gla.ac.uk}{\nolinkurl{m.li.3@research.gla.ac.uk}}} \\
   \And
    Daniela Castro-Camilo
   \\
    School of Mathematics and Statistics, University of Glasgow,
Glasgow, the United Kingdom \\
   \\
  \texttt{\href{mailto:daniela.castrocamilo@glasgow.ac.uk}{\nolinkurl{daniela.castrocamilo@glasgow.ac.uk}}} \\
  }

\newlength{\cslhangindent}
\newlength{\csllabelwidth}
\newlength{\cslentryspacingunit} % times entry-spacing
  {}%
  {\par}
 {% don't indent paragraphs
  \setlength{\parindent}{0pt}
  \ifodd #1
  \let\oldpar\par
  \def\par{\hangindent=\cslhangindent\oldpar}
  \fi
  \setlength{\parskip}{#2\cslentryspacingunit}
 }%
 {}

\definecolor{ml}{RGB}{34,139,34}

\def\gpd{\text{GPD}}
\def\mgpd{\text{mGPD}}
\def\efcm{\text{eFCM}}
\def\hw{\text{HW}}
\def\E{\text{E}}

\begin{document}
\maketitle

\begin{abstract}
Extreme weather events are becoming more frequent and intense, posing serious threats to human life, biodiversity, and ecosystems. 
A key objective of extreme event attribution (EEA) is to assess whether and to what extent anthropogenic climate change influences such events. 
Central to EEA is the accurate statistical characterization of atmospheric extremes, which are inherently multivariate or spatial due to their measurement over high-dimensional grids.
Within the counterfactual causal inference framework of Pearl, we evaluate how tail assumptions affect attribution conclusions by comparing three multivariate modeling approaches for estimating causation metrics. These include: (i) the multivariate generalized Pareto distribution, which imposes an invariant tail dependence structure; (ii) the factor copula model of \cite{castro2020local}, which offers flexible subasymptotic behavior; and (iii) the model of \cite{huser2019modeling}, which smoothly transitions between different forms of extremal dependence. We assess the implications of these modeling choices in both simulated scenarios (under varying forms of model misspecification) and real data applications, using weekly winter maxima over Europe from the Météo-France CNRM model and daily precipitation from the ACCESS-CM2 model over the U.S.
Our findings highlight that tail assumptions critically shape causality metrics in EEA. 
Misspecification of the extremal dependence structure can lead to substantially different and potentially misleading attribution conclusions, underscoring the need for careful model selection and evaluation when quantifying the influence of climate change on extreme events.
\end{abstract}


\section{Introduction}\label{sec:introduction}

In environmental sciences, extreme events are typically defined as instances where a weather or climate variable exceeds a specified threshold. 
Statistically, such events correspond to values in the tails of probability distributions.
As such, their probability of occurrence is small, but their correct estimation is crucial for adequate risk assessment. 
In our current changing climate, it is no surprise that extreme events have become a source of increasing concern due to their potential for substantial impacts on human life and the natural environment. 
In its 2022 Global Risks Report, the World Economic Forum identified extreme weather events, climate inaction, and biodiversity loss as among the most pressing global risks, noting that over half of global GDP depends on nature and urging immediate action to halt nature loss through sustainable practices~\citep{wef2022}.
These concerns are underscored by the increasing intensity and frequency of climate-related extreme events~\citep{mcphillips2018defining}. 
The IPCC’s Sixth Assessment Report further affirms that human-induced greenhouse gas emissions have contributed to a rise in the frequency and/or intensity of certain weather and climate extremes since pre-industrial times~\citep{masson2021climate}.

In recent years, there has been a growing interest in extreme event attribution (EEA), the science of quantifying the influence of human activities on the probability of observed extreme weather or climate events \citep{van2021pathways}. 
Put simply, EEA seeks to determine whether human actions have made an event more likely or more severe. 
In the context of climate change, EEA helps answer questions about the extent to which anthropogenic forcing has altered the probability of particular types of extreme events.
A common approach to EEA involves comparing the probability of an extreme event under the factual scenario, representing observed climatic conditions, with that under a counterfactual scenario in which anthropogenic emissions and land-use changes had not occurred \citep{pearl2009}. 
The counterfactual represents a potential outcome: the state of the world that would have occurred in the absence of the causal factor \citep{shadish2002}.
Formally, let \(Y(1)\) denote the potential outcome under treatment $T=1$ (e.g., presence of human influence), and \(Y(0)\) the potential outcome under treatment $T=0$ (e.g., absence of human influence). 
Estimating the causal effect of $T$ on $Y$ lies at the heart of causal inference, and requires a set of identifying assumptions.
We adopt the Stable Unit Treatment Value Assumption (SUTVA), which includes (i) \emph{consistency}: the observed outcome equals the potential outcome under the treatment actually received, and (ii) \emph{no interference}: one unit’s potential outcomes are unaffected by the treatment status of other units, and treatment is consistently defined across all units. 
Additionally, we assume: (iii) \emph{unconfoundedness}, meaning that $(Y(0),Y(1))$ are independent of $T$ given a set of observed covariates $\pmb{X}$, and (iv) \emph{positivity}, which requires that for all values of $\pmb{X}$, there is a non-zero probability of receiving either treatment (\(0 < P(T = 1 \mid \pmb{X}) < 1\)). Together, these assumptions allow for valid estimation of causal effects from observational data.

In the context of extreme event attribution, let $Y=\mathbb{I}(Z>v)$ denote the binary indicator of an extreme event occurring, where $Z$ is a climatological index and $v$ is a high threshold. 
The treatment $T$ represents the presence ($T=1$) or absence ($T=0$) of anthropogenic forcing. 
We define the probabilities $p_0=P(Y(0)=1)$ and $p_1=P(Y(1)=1)$ as the probabilities of the extreme event under the counterfactual and factual scenarios, respectively.
Within Pearl’s causal framework, various notions of causation—explored in detail in Section~\ref{sec:methods}—can be expressed in terms of $p_0$ and $p_1$, following the work of~\cite{hannart2016}.
It is important to note that these are fundamental quantities, but because we cannot observe repeated realizations of the world, estimates of $p_0$ and $p_1$ typically rely on numerical simulations from climate models under factual and counterfactual scenarios.

Although EEA is a relatively recent area of study, a range of statistical models have been developed to estimate and assess the variability of the probabilities $p_0$ and $p_1$. 
In climatology, non-parametric approaches are commonly employed, modeling the number of threshold exceedances as a binomial count process, where a ``success'' is recorded when the observed value surpasses a predefined threshold. 
Risk ratios are then derived by comparing exceedance counts under factual and counterfactual scenarios~\citep{paciorek2018quantifying}. 
\citet{paciorek2018quantifying} also introduced a formal statistical framework for quantifying sampling uncertainty in risk ratio estimation; however, these methods can become computationally demanding when dealing with rare events defined by high thresholds~\citep{Naveau2020}.
Parametric models based on distributional assumptions provide a useful alternative to non-parametric methods. Though they require selecting a specific distribution, these models generally offer better statistical performance and greater robustness to small estimated probabilities. Extreme-value theory (EVT) is a particularly fitting framework for EEA because it focuses on the stochastic behavior of extreme events in the tails of probability distributions, and it offers tools to assess the probability of events more extreme than any previously observed. However, EVT models have not been widely applied to EEA. In a univariate setting, \cite{van2017rapid} applied the generalized extreme-value (GEV) distribution to annual maxima of 3-day precipitation averages in southern Louisiana. This method estimates the probability of an event with magnitude at least as large as those observed annually, using both observational and model-based data. These estimates are then used in rapid attribution studies, such as for flood-inducing extreme precipitation. Additionally, \cite{gonzalez2025statistical} introduced a streamlined framework for defining non-stationary climate records and quantifying how the likelihood of record-breaking precipitation events changes over time, providing a robust approach to attributing recent extremes to climate trends.
Further, \cite{nanditha2024strong} applied the extended generalized Pareto distribution to attribute daily precipitation extremes across the U.S. to anthropogenic emissions, eliminating arbitrary threshold selection and providing strong evidence of human influence on extreme rainfall events.
However, these approaches typically estimate the desired probabilities location by location, ignoring the natural dependence between sites. This dependence is likely to vary and may differ between the factual and counterfactual worlds. \cite{kiriliouk2020climate} were among the first to explore how multivariate EVT could be applied to EEA. They used the multivariate generalized Pareto distribution (mGPD) to model multivariate extreme behavior when at least one component of the vector is large, combining it with Pearl’s counterfactual causation theory to enhance causal inference in high-dimensional settings.

Following \citet{kiriliouk2020climate}, we focus on high-threshold multivariate exceedance events, defined as cases where at least one component of a multivariate vector exceeds a specified threshold. 
For example, each component might represent precipitation at a different location within a study region, and our interest lies in events where precipitation exceeds a predefined threshold at one or more of these locations. 
Our goal is to assess how assumptions about marginal and joint tail behavior influence the estimation of relevant causal metrics. 
To this end, we compare three distinct modeling approaches.
The first is the multivariate generalized Pareto distribution (mGPD), a widely used, asymptotically justified model for multivariate threshold exceedances. 
The mGPD assumes asymptotic dependence, i.e., that dependence persists even at extreme levels, which may not be appropriate for many environmental applications, where extremal dependence often weakens with increasing severity.
The second approach is the exponential factor copula model (eFCM) of \citet{castro2020local}, a subasymptotic model that allows extremal dependence to decay with event severity. 
It introduces a common latent factor influencing all components, making it particularly suitable for modeling dependence across small spatial domains.
The third model, proposed by \citet{huser2019modeling} (henceforth the HW model), also operates in the subasymptotic regime but offers greater flexibility than either the mGPD or eFCM, accommodating a broader range of extremal dependence structures.
Although originally developed for spatial data, both the eFCM and HW models are adapted here for multivariate settings by constructing appropriate covariance matrices from their underlying covariance functions.

To evaluate the influence of tail dependence and marginal modeling on extreme event attribution, we conduct two simulation studies.
The first explores the impact of model misspecification by deliberately misrepresenting either the marginal distribution or the dependence structure, allowing us to disentangle their respective contributions to errors in tail estimation and, subsequently, errors in causal metrics.
The second compares the performance of the mGPD and eFCM when applied to data generated from the more flexible HW model, which serves as a validation benchmark. 
Across both settings, our results consistently show that, as expected, accurate modeling of extremal dependence has a greater impact on attribution metrics than correct marginal specification. 
The eFCM, in particular, provides the most reliable estimates of extreme event probabilities, effectively capturing the underlying dependence structure.

These simulation results are mirrored in two real-world applications, where we estimate causation metrics, specifically, the attribution ratio (AR) and the probability of necessary causation (PN), using two datasets. 
The first comprises daily precipitation outputs over Europe from the French CNRM global climate model, as studied in \cite{kiriliouk2020climate}, while the second uses daily precipitation data from the ACCESS-CM2 model, part of the CMIP6 ensemble, as described in \cite{nanditha2024strong}.
In both cases, the choice of tail dependence model significantly affects the estimated causation metrics. 
The eFCM and HW models produce dependence structures more consistent with empirical estimates, while the mGPD tends to yield inflated attribution values.
These findings underscore the critical importance of appropriately modeling extremal dependence when drawing causal inferences in extreme event attribution.

The remainder of the paper is structured as follows.
Section~\ref{sec:methods} outlines the methodological framework, introducing the setup for extreme event attribution and describing the three modeling approaches: the multivariate generalized Pareto distribution, the exponential factor copula model, and the Huser–Wadsworth model.
Section~\ref{sec:simu} presents two simulation studies that investigate how marginal specification and dependence structure affect the estimation of attribution ratios.
Section~\ref{sec:app} reports results from two real-world applications based on daily precipitation outputs from the French CNRM global climate model and the ACCESS-CM2 model over the U.S.
Finally, Section~\ref{sec:conclusion} summarizes the key findings and their implications for extreme event attribution.

The C++ and R codes used to implement the three models, replicate the simulation studies, and reproduce the data applications are available at \url{https://github.com/MengranLi-git/tail-eea}.

%%%%%%%%%%%%%%%%%%%%%%%%%%%%%%%%%%%%%%%%%%%%%%%%%%%%%%%%%%%%%%%%%%%%%%%%%%
%%%%%%%%%%%%%%%%%%%%%%%%%%%%%%%%%%%%%%%%%%%%%%%%%%%%%%%%%%%%%%%%%%%%%%%%%%
\section{Methodological framework}\label{sec:methods}

In this section, we define what constitutes an extreme event within the framework of attribution theory. We then present three candidate models, each with distinct capabilities for capturing tail dependence, and explain how they are integrated into the EEA framework.

\subsection{Extreme events attribution}\label{sec:eea}
As outlined in Section~\ref{sec:introduction}, $Y$ represents the binary indicator of an extreme event, assumed to occur when a climatological index exceeds a high threshold $v$.
In the multivariate setting, following \citet{kiriliouk2020climate}, let \(X=(X_1,\cdots,X_d)\top\) be a random vector of climate-related variables at $d$ locations. 
We define the extreme event indicator as
$$Y = \mathbb{I}\{\pmb{w}^\top\pmb{X}>v\},$$
where \(w=(w_1, ..., w_d)^\top\) is a vector of nonnegative weights. 
Let \(\pmb{X}^{(0)}\) and \(\pmb{X}^{(1)}\) denote the climatological vectors under $T=0$ (counterfactual world) and $T=1$ (factual world), respectively. 
Then, the event probabilities can be expressed as
\begin{equation}\label{eq:extremeprobs}
    p_t=\mathbb{P}[\pmb{w}^\top\pmb{X}^{(t)}>v],\quad t\in\{0,1\}.
\end{equation}
To relate $p_0$ and $p_1$ to causal notions, we invoke the \emph{do-operator} to represent hypothetical interventions.
The probability of necessary causation (PN), which quantifies how necessary $T=1$ was for the event, is defined as
\[\mathrm{PN} = \mathbb{P}[ Y(0)=0 \mid \mathrm{do}(T=0), T=1, Y(1)=1] = \max\left(1-\frac{p_0}{p_1},0\right).\]
The probability of sufficient causation (PS), measures how sufficient $T=1$ was for producing the event:
\[\mathrm{PS} = \mathbb{P}[ Y(1) = 1\mid \mathrm{do}(T=1), T=0, Y(0)=0] = \max\left(1-\frac{1-p_0}{1-p_1},0\right).\]
The probability of necessary and sufficient causation (PNS), captures the extent to which the occurrence of the event depends on the treatment being both necessary and sufficient:
\[\mathrm{PNS} = \mathbb{P}[ Y(1) = 1\mid \mathrm{do}(T=1))] -  \mathbb{P}[Y(0)=1 \mid \mathrm{do}(T=0)] = \max\left(p_1-p_0,0\right)\]

These causal measures can be complemented by two commonly used metrics in extreme event attribution: the risk ratio and the attribution ratio. 
The risk ratio (RR) is the factor by which the probability of the event increases due to $T=1$ (anthropogenic influence) and is defined as $$\text{RR} = p_1/p_0.$$
The attribution ratio (AR) represents the probability that the treatment $T=1$ was necessary for the event, given that the event occurred. It is defined as
$$\text{AR} = (p_0-p_1)/p_1.$$ 
Under the standard causal assumptions outlined in Section~\ref{sec:introduction}, all these quantities can be directly estimated in terms of estimates of $p_0$ and $p_1$.
In particular, when $p_0< p_1$, AR is negative, indicating that the presence of anthropogenic influence increased the probability of the event compared to a counterfactual scenario without it.
Conversely, if $p_1 < p_0$, AR becomes positive, suggesting that the event would have been more likely in the absence of the anthropogenic factor. In such cases, PN is zero, reflecting that the treatment was not required for the event to occur.
In this work, we focus on the attribution ratio as a central measure of causal impact, as it captures both risk-enhancing and risk-reducing effects of human activities on extreme climate events. 
Detailed results for PN are reported in the Supplementary Material.

\subsection{Multivariate generalized Pareto distribution}\label{sec:mgpd}

The multivariate generalized Pareto distribution (mGPD) represents the limiting distribution of multivariate exceedances. 
Following \cite{rootzen2006multivariate}, we define a multivariate exceedance as an event where at least one component of the random vector exceeds its corresponding threshold.
To construct a generalized Pareto random vector  $\pmb{Z}$, we follow the approach outlined in \cite{rootzen2018multivariate}.
First, define \[\pmb{Z}^* \overset{d}{=} E+\pmb{T}-\max\limits_{1\leq j\leq d}{T_j},\] where \(E\) is a unit exponential random variable and \(\pmb{T}=(T_1, \cdots, T_d)^\top\) represents any d-dimensional random vector independent of \(E\). 
Then, for \(\pmb{\sigma} = (\sigma_1,\ldots,\sigma_d)^\top>\pmb{0}\) and \(\pmb\gamma = (\gamma_1,\ldots,\gamma_d)^\top\in\mathbb{R}^d\), the generalized Pareto random vector $\pmb{Z}$ is given by
\[\pmb{Z}\overset{d}{=}\frac{\pmb{\sigma}}{\pmb{\gamma}}\left(\exp(\pmb{\gamma Z}^*)-1\right),\] where operations like \(\pmb{\frac{\sigma}{\gamma}}\) are performed componentwise. 
We denote the distribution of $\pmb{Z}$ as \(\pmb{Z}\sim \mgpd(\pmb{T},\pmb{\sigma},\pmb{\gamma})\).
If $\pmb{T}$ has density $f_{\pmb{T}}$, the density of $\pmb{Z}^*$ is given by
\begin{equation}\label{eq:mgpd_densityunitz}
    f_{\pmb{Z}^*}(\pmb{z}) = \frac{1\{\max(\pmb{z}) >0\}}{\exp\{\max(\pmb{z})\}}\int_0^\infty f_{\pmb{T}}(\pmb{z} + \log t)t^{-1}\text{d}t,
\end{equation}
and the density of $\pmb{Z}$ can be expressed as
\begin{equation*}
    f_{\pmb{Z}}(\pmb{z}) = f_{\pmb{Z}^*}\left(\frac{\mathds{1}}{\pmb{\gamma}}\log(1 + \pmb{\gamma}\pmb{z}/\pmb{\sigma})\right)\prod_{j=1}^d\frac{1}{\sigma_j + \gamma_j z_j}.
\end{equation*}
Parametric models for $\pmb{Z}$ are thus constructed by specifying a distribution for the latent vector $\pmb{T}$, referred to as the generator.
\cite{kiriliouk2019peaks} derived explicit densities for different types of generators.
Here, we focus on a class of generators with independent reverse exponential components. Specifically, we assume that $f_{\pmb{T}}(\pmb{z}) = \prod_{j=1}^df_j(z_j)$ where $f_j(z_j) = \alpha_j\exp\{\alpha_j(z_j+\beta_j)\}$, with $z_j<-\beta_j$, $\alpha_j>0$, $\beta_j\in\mathbb{R}$.
Under this assumption, the density in~\eqref{eq:mgpd_densityunitz} becomes
\begin{equation*}
    f_{\pmb{Z}^*}(\pmb{z}) = \frac{\exp\left\{-\max(\pmb{z}) - \max(\pmb{z} + \boldsymbol{\beta})\sum_{j=1}^d\alpha_j\right\}}{\sum_{j=1}^d\alpha_j}
    \prod_{j=1}^d\alpha_j(\exp\left\{\alpha_j(z_j+\beta_j)\right\}).
\end{equation*}
Although the marginal distributions of the mGPD are not necessarily univariate GPDs, this result holds when considering conditional marginal distributions.
Specifically, the $j$-th conditional marginal survival function corresponds to the survival function of the univariate GPD $H$, such that
\begin{equation}\label{eq:marginalGPD}
    \mathbb{P}[Z_j>z\mid Z_j>0]=\bar H(z;\sigma_j,\gamma_j),
\end{equation}
 for any \(z>0\) and \(j\in\{1, ...,d\}\), where
 \begin{equation}\label{eq:survivalgpd}
     \bar H(z;\sigma_j,\gamma_j)=\left(1+\gamma \frac{z}{\sigma}\right)_+^{-1/\gamma}, \quad \sigma>0,\ \gamma\in\mathbb{R}.
 \end{equation}
Assuming that $\gamma_j\equiv\gamma$, i.e., the shape parameter is the same for all locations, \eqref{eq:marginalGPD} can be computed using linear projections, as described in~\cite{rootzen2018multivariate}.
Specifically, if \(\pmb{Z}\sim \mgpd(\pmb{T},\pmb{\sigma},\gamma\pmb{1})\), then for any nonnegative weights
\(\pmb{w}=(w_1, \cdots, w_d)^\top\) such that \(P(\pmb w^\top \pmb Z>0)>0\), the conditional distribution of $\pmb w^\top \pmb Z$given $\pmb w^\top \pmb Z>0$ is
\begin{equation}\label{eq:gpdproj}
    [\pmb w^\top \pmb Z\mid \pmb w^\top \pmb Z>0]\sim \gpd(\pmb w^\top\pmb\sigma,\gamma).
\end{equation}
Assuming tail homogeneity across the study region requires prior analysis to assess its suitability in practice. 
While the result in~\eqref{eq:gpdproj} offers computational efficiency for marginal probability estimation, the mGPD model remains computationally demanding overall.
Additionally, its underlying assumption of persistent strong tail dependence at increasingly extreme levels may not hold in many environmental applications, where evidence often points to weakening dependence in the tails. 
These considerations motivate the use of the two alternative subasymptotic models introduced below.

\subsection{Exponential factor copula model}\label{sec:efcm}

\cite{castro2020local} propose a censored local likelihood approach for a subasymptotic spatial process $W$, stochastically defined as the sum of a Gaussian process and a common exponential factor.
Specifically
\begin{equation}\label{eq:efcm}
    W(s)=Z(s)+V, \quad s\in \pmb{S}\subset \mathbb{R}^2,
\end{equation}
where $\pmb{S}$ denotes the study area, $Z(s)$ is a zero-mean Gaussian process with a stationary correlation function \(\rho(h)\) (with \(h=\mid \mid s_1-s_2\mid \mid ,s_1,s_2\in\pmb{S}\)), and \(V\geq0\) is an exponentially distributed random variable with rate parameter \(\lambda>0\), independent of both \(Z(s)\) the and spatial location \(s\in \pmb{S}\). 
Thus, the process \(W\), which falls within the class of asymptotically dependent models, can be viewed as a Gaussian location mixture, i.e., a Gaussian process with a random (exponentially distributed) mean. 
Letting \(Z_j=Z(s_j),j=1,...,d\), the random vector \(\pmb{Z}=(Z_1, ...,Z_d)^\top\) follows a multivariate normal distribution, \(\pmb{Z}\sim \Phi(\cdot;\Sigma_Z)\), where the covariance matrix \(\Sigma_Z\) is determined by the correlation function \(\rho(h)\).
Here, we specify  \(\rho(h)\) as the exponential correlation function \(\rho(h)=\exp\{-h/\delta\}\), with \(\delta\) denoting the range parameter.
We similarly define  \(\pmb{W}=(W_1, ...,W_d)^\top\), where \(W_j = Z_j+V\), for $j=1,...,d$, and \(V\sim \exp(\lambda)\) is independent of \(\pmb Z\).

The inclusion of the common factor $V$ restricts the model in~\eqref{eq:efcm} to homogeneous regions, akin to the tail homogeneity assumption underlying the result in~\eqref{eq:gpdproj}. 
As a result, preliminary checks for homogeneity across locations are necessary before applying the model.
The censored local likelihood approach proposed by \cite{castro2020local} addresses this limitation by partitioning the study area into smaller, approximately homogeneous regions, allowing the model to be applied more flexibly across larger and more heterogeneous spatial domains.

\subsection{Huser and Wadsworth model}\label{sec:hw}
\cite{huser2019modeling} propose a flexible stationary model that captures both asymptotic dependence and asymptotic independence in spatial extremes, allowing a smooth transition between these two classes without requiring pre-specification. The HW model is defined as
\begin{equation}\label{eq:hw}
    X(\pmb s)=R^{\delta}W(\pmb s)^{1-\delta}, \quad \delta\in[0,1],\ \pmb s\in \pmb{S}\subset \mathbb{R}^2,
\end{equation}
where \(W(\pmb s)\) is a stationary spatial process with standard Pareto margins, $R$ is an independent standard Pareto random variable, and $\delta$ controls the strength of extremal dependence. 
When $\delta>1/2$, $R^{\delta}$ dominates, leading to asymptotic dependence; when $\delta<1/2$,  $W(\pmb s)^{1-\delta}$ dominates, resulting in asymptotic independence.
Inference is performed using a censored likelihood approach based on a copula construction. The choice of model for \(W(\pmb s)\) can vary depending on the application, with the authors exploring options such as Gaussian processes, inverted extreme value logistic model and inverted max-stable processes.
Here, we will use Gaussian processes to model the dependence structure.
As with the model in~\eqref{eq:efcm}, the HW model can be adapted to the multivariate setting by replacing the process $W(s)$ with an asymptotically independent random vector $(W_1,\ldots,W_d)^\top$.

\subsection{Dependence coefficients}\label{sec:dependence}
To understand how different models represent joint extremes, it is important to examine the extremal dependence structures they imply. 
This can provide insight into potential differences in estimated causal probabilities, as introduced in Section~\ref{sec:eea}. 
In this section, we characterize extremal dependence using two measures: the bivariate tail dependence coefficient $\chi\in[0,1]$ and the conditional exceedance probability $\chi(u)$.
The coefficient $\chi$ captures the strength of dependence in the asymptotic (limiting) tail, while $\chi(u)$ provides a finite-level analogue, measuring the probability of joint exceedances at high but non-asymptotic thresholds.
For vector components $Y_1$ and $Y_2$, these coefficients are defined as 
\[\chi= \lim_{u\to1}\chi(u):= \lim_{u\to1}\Pr(F_{Y_1}(Y_1)>u\mid F_{Y_2}(Y_2)>u).\]

For the multivariate generalized Pareto distribution (mGPD) with independent reverse exponential generators, the bivariate tail dependence coefficient $\chi_\mgpd$ admits a closed-form expression (see the Supplementary Material of~\citealt{kiriliouk2019peaks}):
$$\chi_\mgpd = 1 - \left(\frac{1 - \alpha_{(1)}^{-1}}{1 + \alpha_{(2)}^{-1}}\right)^{1 + \alpha_{(2)}}\frac{\alpha_{(1)}}{\alpha_{(2)}}\frac{1}{1 + \alpha_1 + \alpha_2},$$
where $\alpha_{(1)}  = \max(\alpha_1,\alpha_2)$ and $\alpha_{(2)}  = \min(\alpha_1,\alpha_2)$.
The corresponding $\chi_\mgpd(u)$ does not have a closed-form expression for finite $u$ but can be estimated numerically, e.g., via Monte Carlo or quadrature.

For the eFCM, the tail dependence coefficient is given by
$$\chi_\efcm = 2\left\{1-\Phi\left(\lambda\sqrt{\frac{(1 -\rho_{12})}{2}}\right)\right\},$$
where $\lambda$ is the rate parameter and $\rho_{12}$ is the correlation (in our case, derived from an exponential correlation model).
The corresponding $\chi_\efcm(u)$ can be computed as
$$\chi_\efcm(u) = \frac{1 - 2u + \Phi_2(z(u), z(u); \rho)
- 2\exp\{\lambda^2/2 - \lambda z(u)\Phi_2(q; 0, \Omega)\}}{1 - u},
$$
where
$z(u) = F_1^{-1}(u;\lambda)$ is the marginal quantile function, $\Phi_2(\cdot,\cdot,
\rho)$ is the bivariate standard normal CDF with correlation $\rho$, $q = \lambda(1-\rho)$, and $\Omega$ is the correlation matrix.

For the HW model, closed-form expressions for $\chi_\mgpd$ and $\chi_\mgpd(u)$ are generally not available, but both can be numerically approximated.
If $\delta>1/2$, the process is asymptotically dependent, and the limiting coefficient is
$$\chi_\hw = \E\left\{\min(W_1,W_2)^{(1-\delta)/\delta}\right\}\frac{2\delta-1}{\delta}$$
Moreover, if $\delta\neq 1/2$  and $u$ is near 1, we have
$$\chi_\hw(u) = \frac{1 - 2u + \displaystyle \int_0^{\log w(u) / \delta} F_{\tilde{W}_1, \tilde{W}_2} \left( \frac{\log w(u) - \delta r}{1 - \delta}, \frac{\log w(u) - \delta r}{1 - \delta} \right) e^{-r} \, \text{d}r }{1 - u},$$
where
$w(u) = F_1^{-1}(u)$, is the marginal quantile function of the HW model, and $F_{\tilde{W}_1, \tilde{W}_2}$ is the joint distribution of the log-transformed components $\tilde{W}_1 = \log W_1$ and $\tilde{W}_2 = \log W_2$.
At the boundary case $\delta=1/2$, the process is asymptotically independent, so $\chi_\hw = 0$.
Nevertheless, the decay of $\chi_\hw(u)$ is unusually slow:
$$\chi_\hw(u) \sim \frac{1-u}{\log\{1/(1-u)\}},\quad \text{as } u\to 1,$$
i.e., $\chi_\hw(u)$ decays to zero more slowly than any power of $1-u$.
Step-by-step derivations of $\chi_\efcm(u)$ and $\chi_\hw(u)$ are provided in Section~\ref{sec:appendix_chi} of the Appendix.

\subsection{Computation of extremal probabilities $p_0$ and $p_1$}\label{sec:extremeprobs}
In this section, we describe how to compute the extreme probabilities in~\eqref{eq:extremeprobs} using the three models introduced in Sections~\ref{sec:mgpd}–\ref{sec:hw}. We also explain the construction of the nonnegative weight vector \(\pmb{w}=(w_1, \cdots, w_d)^\top\) used in these calculations.

Let \(\pmb{u} = (u_1, \cdots, u_d)^\top\) be a vector of high marginal thresholds, and let \(\pmb{w}=(w_1, \cdots, w_d)^\top\) be the vector of nonnegative weights such that $v>\pmb{w}^\top\pmb{u}$.

Under the mGPD model, and using the projection property in~\eqref{eq:gpdproj}, the probability of a weighted exceedance can be approximated by
\begin{equation}\label{eq:extprob_mgpd}
\begin{aligned}
\mathbb{P}[\pmb{w}^\top\pmb{X}>v]
&=\mathbb{P}[\pmb{w}^\top\pmb{X}>\pmb{w}^\top\pmb{u}]\cdot\mathbb{P}[\pmb{w}^\top(\pmb{X}-\pmb{u})>v-\pmb{w}^\top\pmb{u}\mid \pmb{w}^\top(\pmb{X}-\pmb{u})>0]\\
& \approx \mathbb{P}[\pmb{w}^\top\pmb{X}>\pmb{w}^\top\pmb{u}]\cdot\bar{H}(v-\pmb{w}^\top\pmb{u};\pmb{w}^\top\pmb{\sigma}, \gamma),
\end{aligned}
\end{equation}

where $\bar{H}(\cdot;\pmb{w}^\top\pmb{\sigma}, \gamma)$ is the survival function of the univariate GPD with scale parameter $\pmb{w}^\top\pmb{\sigma}$ and common shape parameter $\gamma$ (see~\eqref{eq:survivalgpd}).
The exceedance probability in the first term of~\eqref{eq:extprob_mgpd} can be estimated empirically as
$$\widehat p^{\text{emp}}(\pmb{w}^\top\pmb{u}^{(t)};\pmb w)=\frac{1}{n}\sum_{i=1}^n\mathds{1}\{\pmb{w}^\top\pmb X_i^{(t)}>\pmb{w}^\top\pmb{u}^{(t)}\},\quad t\in\{0,1\}.$$

If $v\leq\pmb{w}^\top\pmb{u}$, then the event lies within the bulk of the distribution, and we estimate the probability $\mathbb{P}[\pmb{w}^\top\pmb{X}>v]$ directly
$$\widehat p^{\text{emp}}(v;\pmb w)=\frac{1}{n}\sum_{i=1}^n\mathds{1}\{\pmb{w}^\top\pmb X_i^{(t)}>v\},\quad t\in\{0,1\}$$
Putting both cases together, we define the mGPD estimator for $p_t = \mathbb{P}[\pmb{w}^\top\pmb{X}^{(t)}>v]$, $t\in\{0,1\}$ as
\[
\widehat p_t(v;\pmb{w})=
\begin{cases}\widehat p_t^{\text{emp}}(v;\pmb w), & \text{if $v\leq\pmb{w}^\top\pmb u^{(t)}$}, \\
\widehat p_t^{\text{emp}}(\pmb{w}^\top\pmb u^{(t)};\pmb{w})\bar H[v-\pmb{w}^\top\pmb u^{(t)};\pmb{w}^\top\hat{\pmb{\sigma}},\hat\gamma], & \text{if $v>\pmb{w}^\top\pmb u^{(t)}$}.
\end{cases}\]

For the eFCM and HW models, the probabilities \(p_t\) are estimated by Monte Carlo integration over samples \(\pmb{X}_i'^{(t)}\) drawn from the fitted models:
\[\widehat p_t(v;\pmb w)=\frac{1}{n}\sum_{i=1}^n\mathds{1}\{\pmb{w}^\top\pmb X_i'^{(t)}>v\},\quad t\in\{0,1\}.\]

To construct the nonnegative weight vector $\pmb{w} = (w_1, \ldots, w_d)^\top$, we follow the approach of~\cite{kiriliouk2020climate}, which proposes maximizing measures of causal attribution, specifically, the probability of necessary causation or the attribution ratio. 
The key idea is to optimize over linear projections $\pmb{w}^\top \pmb{X}$ to identify the direction in the multivariate space that is most sensitive to the influence of external forcing. 
This projection can be interpreted as a near-sufficient statistic for distinguishing factual and counterfactual extremes. 
It enhances statistical efficiency by concentrating on the features that most strongly capture changes in tail probabilities due to forcing. 
Instead of fixing $\pmb{w}$ a priori, the data-driven choice of $\pmb{w}$ allows us to adaptively focus on the dimensions of the system most affected by anthropogenic influence.
To this end, we treat the extreme probabilities as a function of $\pmb{w}$, i.e.,
$$p_t(\pmb{w}) := \mathbb{P}(\pmb{w}^\top X^{(t)} > v).$$
Our goal is to find the weight vector \( \pmb{w}^* \) that maximizes either the PN or AR. 
These are defined as:
$$\pmb{w}^* = \arg\max_{\pmb{w}} \mathrm{PN}(\pmb{w}) := \arg\max_{\pmb{w}} \left(1 - \frac{p_0(\pmb{w})}{p_1(\pmb{w})}\right)_+$$
or
$$\pmb{w}^* = \arg\max_{\pmb{w}}  \mathrm{AR}(\pmb{w}) := \arg\max_{\pmb{w}} \frac{p_0(\pmb{w}) - p_1(\pmb{w})}{p_1(\pmb{w})}$$
The motivation for defining $\pmb{w}^*$ as the maximizer of PN or AR stems from the goal of identifying the most informative linear combination of variables that reveals the effect of anthropogenic forcing on the probability of extreme events. Specifically, this approach: (i) identifies the linear projection most influenced by the forcing; (ii) maximizes contrast between factual and counterfactual exceedance probabilities; (iii) aligns with the counterfactual framework for event attribution; and (iv) supports interpretable attribution statements about which features are driving the observed changes.

%%%%%%%%%%%%%%%%%%%%%%%%%%%%%%%%%%%%%%%%%%%%%%%%%%%%%%%%%%%%%%%%%%%%%%%%%%
%%%%%%%%%%%%%%%%%%%%%%%%%%%%%%%%%%%%%%%%%%%%%%%%%%%%%%%%%%%%%%%%%%%%%%%%%%
\section{Simulation study}\label{sec:simu}
To investigate the impact of tail assumptions on extreme event attribution, we conduct two simulation studies comparing the three multivariate extreme value models introduced in Section~\ref{sec:methods}. The first scenario demonstrates that accurate modeling of extremal dependence plays a more critical role in reliable EEA than precise specification of marginal distributions. The second scenario uses the HW model to simulate data under different asymptotic regimes, evaluating the performance of the mGPD and eFCM models. Results highlight that the enhanced flexibility of the eFCM in capturing subasymptotic dependence leads to more accurate attribution estimates.

\subsection{Simulation scenario 1}\label{sec:simu1}
Accurate specification of both the marginal distributions and the dependence structure is essential in multivariate modeling. 
However, in practice, it is often difficult to get both right—a challenge we also encountered in our data applications. 
In the context of multivariate extreme value analysis, it is widely acknowledged that the dependence structure often plays a more critical role than the marginal distributions in determining joint tail probabilities, particularly when assessing the likelihood of co-occurring extremes.
Here, we aim to quantify the extent to which marginal and dependence components affect tail probability estimation.

We chose the HW model as the true data-generating process because its stochastic representation naturally separates the marginal distributions from the dependence structure.
We generate $n = 1,000$ replicates of the HW model at $d=2$ locations, assuming a Gaussian process for $W$ with an {exponential covariance function} with range parameter $0.8$. 
We consider two values of $\delta$, namely $\delta = 0.3$ (asymptotic independence) and $\delta = 0.7$ (asymptotic dependence).
Let $(\pmb{X}_{1}, \pmb{X}_{2})$ denote the bivariate simulated vectors, with $\pmb{X}_{i} = (X_{i1}, \ldots, X_{in})^\top$, $i = 1, 2$.
Our target is to estimate $p = \Pr(w_1X_1 + w_2X_2 > \nu_p)$, where $\nu_p$ is the $p$-quantile of the weighted sum $w_1X_1 + w_2X_2$, for $p = 0.01, 0.02, 0.05$ and $w_1=w_2=1/2$.  
We repeat the entire simulation procedure $\text{MC} = 1,000$ times to assess model performance. 
To isolate the influence of marginal and dependence misspecification, we consider the two sub-scenarios described below.

\subsubsection{Sub-scenario 1.1: Misspecified margins, correct dependence}
In this setting, we assume the correct dependence structure (``right dependence'' or RD) for $W$ but intentionally misspecify the margins (``wrong margins'' or WM).
Specifically, we introduce marginal distortion by applying the transformation ${\pmb{X}}_{i}^e = F_p^{-1}\{F_e(\pmb{X}_{i})\}$, where $F_e$ is the unit exponential distribution and $F_p^{-1}$ is the inverse of the standard Pareto distribution.
We then fit the HW model to the transformed data  $({\pmb{X}}_{1}^e, {\pmb{X}}_{2}^e)$, using the correct dependence structure (i.e., the original Gaussian process with true covariance). 
From the fitted model, we generate $m=100,000$ samples $(\tilde{X}_{1,1},\ldots,\tilde{X}_{1,m},\tilde{X}_{2,1},\ldots,\tilde{X}_{2,m})^\top$ and estimate $p$ using the empirical estimator $\hat{p} = \frac{1}{m}\sum_{k=1}^m\mathds{1}\{w_1\tilde{X}_{1,k}+w_2\tilde{X}_{2,k}>\nu_p\}$ for each value of $p$.

\subsubsection{Sub-scenario 1.2: Misspecified dependence, correct margins }\label{sec:subsce2} 
\begin{figure}[!ht]\centering
  \centering
  \includegraphics[width=0.5\linewidth]{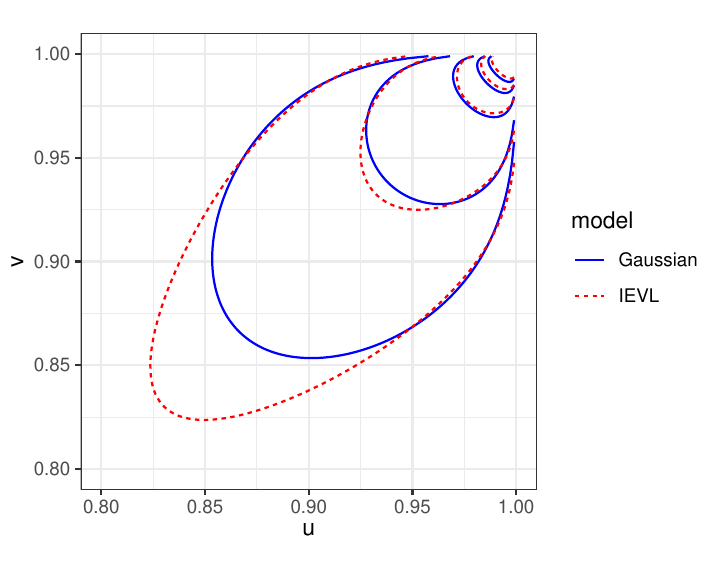}
  \caption{Density contours of the Gaussian and inverted extreme value logistic (IEVL) copulae, plotted over the range $u,v \in [0.80, 0.999]$, related to the simulation sub-scenario 1.2. The density is concentrated near the upper-right corner $(1,1)$ for both copulas, illustrating their upper tail dependence structure. The Gaussian copula exhibits symmetric dependence with weaker tail dependence, while the IEVL copula shows stronger upper tail clustering.}
  \label{fig:simu1_copula}
\end{figure}

Here, the margins are correctly specified as standard Pareto, but we misspecify the dependence structure by using the inverted extreme value logistic (IEVL) model for $W$, which results in incorrect spatial dependence modeling; see Figure \ref{fig:simu1_copula} for a comparison of the Gaussian and IEVL copulae. 
As in Sub-scenario 1.1, we use the fitted model to simulate $m = 100,000$ replicates and estimate
$p$ via its empirical estimator.

\begin{figure}[!ht]\centering
  \centering
  \includegraphics[width=0.7\linewidth]{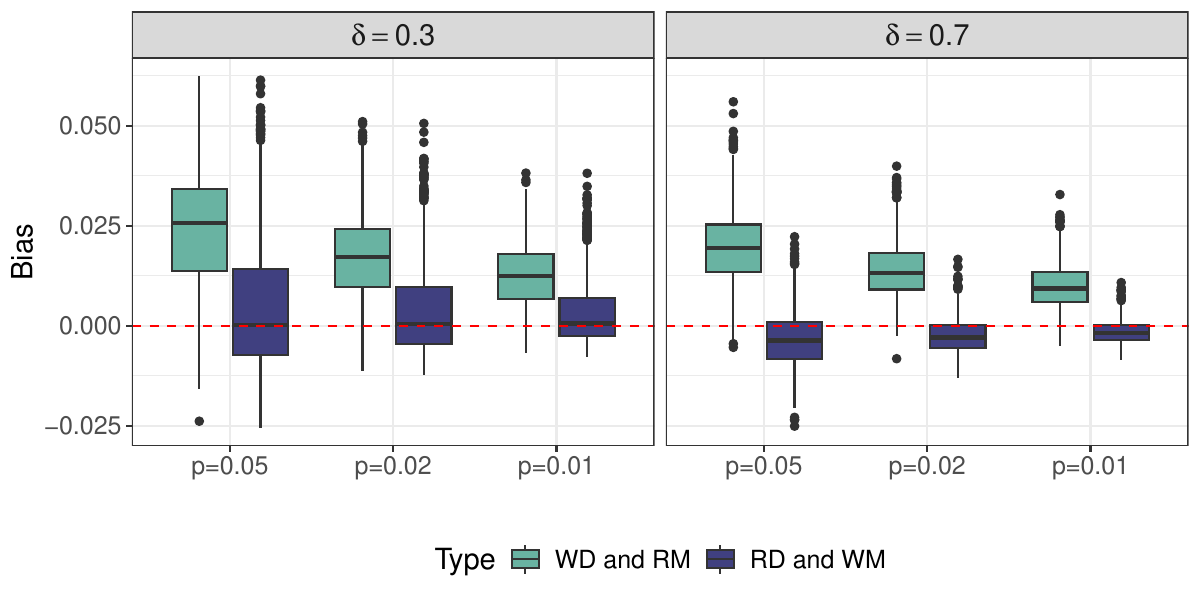}
  \caption{Bias of $\hat{p}$ for $p=0.01, 0.02, 0.05$  with $\delta=0.3$ (left) and $\delta=0.7$ (right) for the two simulation sub-scenarios detailed in Section~\ref{sec:simu1}. The purple boxplot corresponds to sub-scenario 1.1 with right dependence and wrong margins (RD and WM) while the green boxplot corresponds to wrong dependence and right margins (WD and RM).}
  \label{fig:simu1_boxplot}
\end{figure}

Figure~\ref{fig:simu1_boxplot} shows the {bias} of $\hat{p}$ for both sub-scenarios across the different values of $p$, using boxplots based on $\text{MC} = 1,000$ Monte Carlo replicates. 
Purple boxplots correspond to Sub-scenario 1.1 (right dependence RD, wrong margins WM), and green boxplots correspond to Sub-scenario 1.2 (wrong dependence WD, right margins RM). 
The results indicate that while marginal misspecification introduces noise, it preserves the rank structure of the data and has minimal impact on tail probability estimates. In contrast, misspecifying the dependence structure leads to substantial bias, underscoring the greater importance of capturing extremal dependence accurately when performing EEA.

\subsection{Simulation scenario 2}\label{sec:simu2}
Having established the critical role of accurately modeling dependence, we now compare the performance of the mGPD and eFCM models in estimating the tail probabilities $p = \Pr(w_1X_1 + w_2X_2 > \nu_p)$. 
As in the first simulation, we use the HW model as the data-generating process due to its flexibility in capturing various asymptotic dependence regimes.

We generate $n = 1,000$ replicates from the HW model, again assuming a Gaussian process for $W$ with an exponential covariance function and range parameter $0.8$. 
To explore a broader spectrum of asymptotic dependence, we consider five values of the dependence parameter: $\delta = 0.3,\ 0.4,\ 0.6,\ 0.7,\ 0.8$. 
For each setting, we repeat the simulation procedure $\text{MC} = 1{,}000$ times to assess model performance.

\begin{figure}[!ht]
  \centering
  \includegraphics[width=0.7\linewidth]{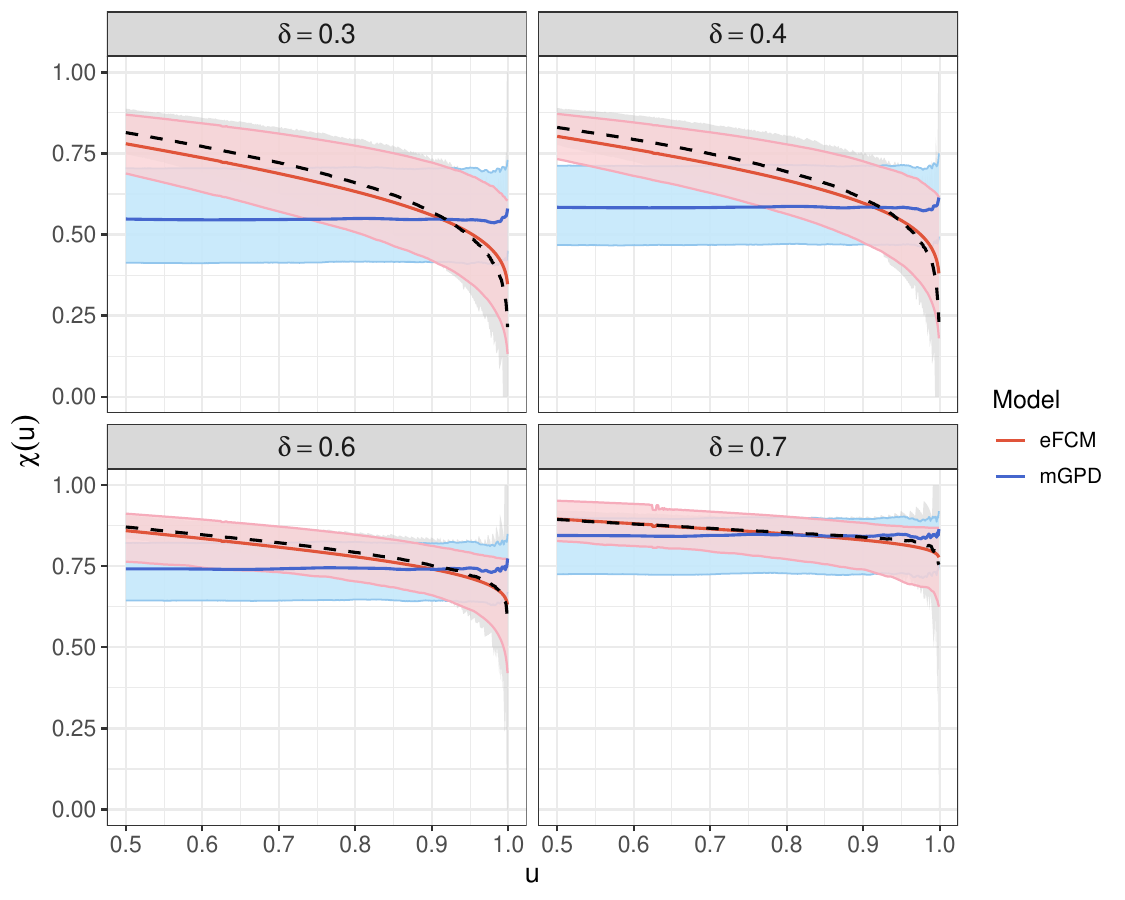}
  \caption{Estimated and true values of $\chi(u)$ for the second simulation study described in Section~\ref{sec:simu2} for $u\in[0.5,1]$ and $\delta = 0.3,\ 0.4,\ 0.6,\ 0.7$. The black line shows the true $\chi(u)$ from the HW model, while the red and blue lines represent estimates from the eFCM and mGPD models, respectively. Shaded regions indicate 95\% pointwise confidence intervals.}
  \label{fig:simu2_chiplot}
\end{figure}

Figure~\ref{fig:simu2_chiplot} shows estimates of the conditional probability $\chi(u)$ for $u \in [0.5, 1]$ obtained from both models, with the true $\chi(u)$ curve from the HW model included for comparison. 
The results for $\delta=0.8$ were omitted as they are almost identical to those of $\delta=0.7$.
As expected, the mGPD displays its characteristic constant $\chi(u)$ profile, which leads to overestimation of extremal dependence ($u$ close to 1) when $\delta < 0.5$ (cases of asymptotic independence) and fails to reflect the decay in dependence typical of subasymptotic regimes. 
In contrast, the eFCM provides a much closer match to the true $\chi(u)$ across all $\delta$ values, successfully capturing the changing dependence structure.

\begin{figure}[!ht]\centering
  \centering
  \includegraphics[width=\linewidth]{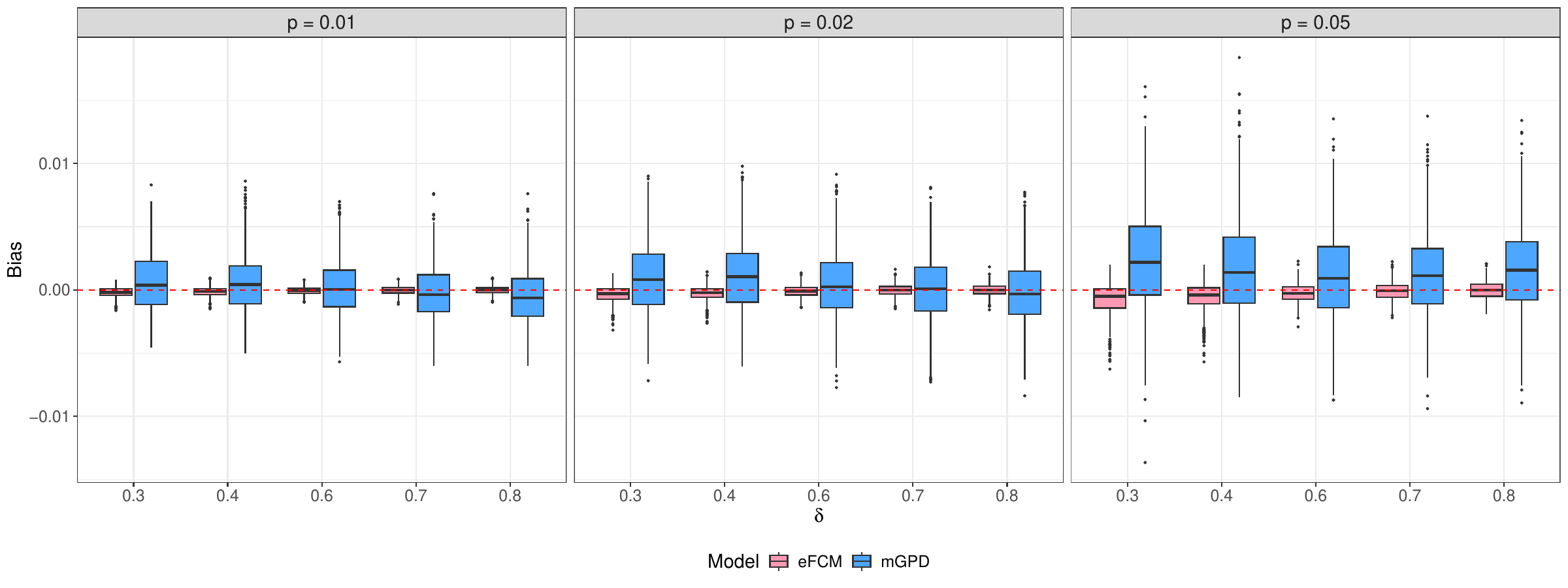}
  \caption{Bias of $\hat{p}$ in the second simulation scenario (Section~\ref{sec:simu2}) using the mGPD and eFCM models for $p=0.005,\ 0.01, \ 0.02$ and $\delta = 0.3,\ 0.4,\ 0.6,\ 0.7,\ 0.8$. Boxplots were constructed using $\text{MC} = 1,000$ Monte Carlo replicates.}
  \label{fig:simu2_boxplot}
\end{figure}

Figure~\ref{fig:simu2_boxplot} presents estimates of the tail probabilities $p = \ 0.01,\ 0.02,\ 0.05$ for each value of $\delta$. 
The greater flexibility of the eFCM in modeling dependence translates into more accurate and consistent probability estimates, especially in settings with weaker asymptotic dependence. 
By comparison, the mGPD shows greater bias and variability, particularly when $\delta$ is small, which is expected given its rigid dependence structure.
Table~\ref{tab:rmse} reports the root mean square error (RMSE) for estimates of $\hat{p}$. 
All in all, the eFCM consistently outperforms the mGPD in all asymptotic dependence regimes.

\begin{table}[ht]
\centering
\caption{Root mean square error (RMSE) of $\hat{p}$ in the second simulation scenario (Section~\ref{sec:simu2}). Tail probabilities $\hat{p}$ are estimated using the mGPD and eFCM models for $p=0.01,\ 0.02,\ 0.05$ and $\delta = 0.3,\ 0.4,\ 0.6,\ 0.7,\ 0.8$. For each combination of $p$ and $\delta$, the model with the lowest RMSE is highlighted in bold.}
\label{tab:rmse}
\begin{tabular}{l c|ccccc}
  & & \multicolumn{5}{c}{$\delta$} \\
  Model & $p$ & 0.3 & 0.4 & 0.6 & 0.7 & 0.8 \\
  \toprule
  eFCM & 0.05 & \textbf{0.0017} & \textbf{0.0012} & \textbf{0.0008} & \textbf{0.0007} & \textbf{0.0007} \\
  mGPD &      & 0.0049 & 0.0043 & 0.0036 & 0.0035 & 0.0038 \\
  \midrule
  eFCM & 0.02 & \textbf{0.0008} & \textbf{0.0006} & \textbf{0.0005} & \textbf{0.0004} & \textbf{0.0004} \\
  mGPD &      & 0.0030 & 0.0031 & 0.0027 & 0.0026 & 0.0025 \\
  \midrule
  eFCM & 0.01 & \textbf{0.0005} & \textbf{0.0004} & \textbf{0.0003} & \textbf{0.0003} & \textbf{0.0003} \\
  mGPD &      & 0.0025 & 0.0024 & 0.0021 & 0.0022 & 0.0022 \\
  \bottomrule
\end{tabular}
\end{table}

%%%%%%%%%%%%%%%%%%%%%%%%%%%%%%%%%%%%%%%%%%%%%%%%%%%%%%%%%%%%%%%%%%%%%%%%%%
%%%%%%%%%%%%%%%%%%%%%%%%%%%%%%%%%%%%%%%%%%%%%%%%%%%%%%%%%%%%%%%%%%%%%%%%%%
\section{Data applications}\label{sec:app}

To assess the extent to which human activities have contributed to extreme precipitation events, we combine the causal inference framework for event attribution (Section~\ref{sec:eea}) with the extreme value modeling techniques introduced in Sections~\ref{sec:mgpd} to~\ref{sec:hw}. 
A central challenge in this setting is the inherent unobservability of the counterfactual world. 
To circumvent this, we use climate model simulations to generate data representing both the factual and counterfactual scenarios.

We focus on two applications concerning precipitation extremes. 
The first dataset contains simulations from the CNRM global climate model developed by Météo-France, part of the Coupled Model Intercomparison Project Phase 6 (CMIP6; \citealt{eyring2016overview,voldoire2019evaluation}), providing precipitation outputs over Europe under both factual and counterfactual conditions. 
The second dataset offers similar simulations across the contiguous United States, based on the ACCESS-CM2 model (which also contributed to CMIP6; \citealt{bi2013access}). 
Both datasets have been studied previously in the context of EEA in~\cite{kiriliouk2020climate} and~\cite{nanditha2024strong}, respectively.
Our aim is to evaluate how varying modeling assumptions influence the assessment of anthropogenic forcing on the occurrence of extreme precipitation events across these regions.

The spatial nature of both datasets means that the Stable Unit Treatment Value Assumption (SUTVA) does not hold. 
Spatial dependence among locations violates the assumption of unit-level independence, while spatial heterogeneity undermines the premise of a uniform treatment effect.
Additionally, internal climate variability may mask or distort the true impact of the treatment, leading to potentially misleading inferences. 
These challenges are not unique to our study but are common to spatial applications of causal inference.
To mitigate these violations in the context of event attribution, we first reduce heterogeneity by clustering locations with similar features. 
Within each cluster, where spatial dependence may still be present, we fit the three models introduced in Section~\ref{sec:methods}, each designed to account for dependence structures in both the factual and counterfactual worlds. 
We then estimate and compare the attribution ratio (AR) across scenarios to assess the influence of human activity on precipitation extremes. This comparison relies on accurate estimation of extreme probabilities, as described in Section~\ref{sec:extremeprobs}.
To estimate these probabilities, we define appropriate thresholds in both worlds. 
Specifically, each component $u_j$ of the vector $\pmb{u}$ is set to the 90th percentile of the empirical distribution at location $j$, computed separately for the factual and counterfactual scenarios. The scalar $v$ is chosen to correspond to the 5-year and 50-year return levels under the counterfactual climate.
Estimation results for the probability of necessary causation (PN) across all models and both applications are presented in Appendix Section~\ref{sec:appendix_pn}.
 
\subsection{Precipitation in Europe using CNRM outputs}\label{sec:app1}

The CNRM dataset contains daily maximum precipitation values for the winter months (1\textsuperscript{st} January 1985 to 31\textsuperscript{st} December 2014; 2,707 time points) across 433 locations in central Europe.
The factual and counterfactual scenarios are represented by two historical simulations, one of which includes only natural forcings. 
To reduce potential nonstationarity over time, we follow \citet{kiriliouk2020climate} and compute the 7-day maxima at each location, treating these as temporally stationary series within each world.

As outlined in Section~\ref{sec:methods}, the three models under consideration assume tail homogeneity across the spatial domain. To satisfy this assumption, we follow the approach of \citet{castro2020local} and partition the study region into homogeneous clusters, defined as regions where the marginal distributions of threshold exceedances are approximately stationary. 
Each model is then fitted locally within these clusters.

To define the homogeneous clusters, we begin by constructing a regular grid of 881 centroids, spaced 1 degrees apart in both latitude and longitude. For each grid point $\mathbf{s}_0$, we define a cluster $\mathcal{N}_{\mathbf{s}_0, D_0}$ consisting of the $D_0$ nearest CNRM model output locations. 
The value of $D_0$ is chosen to ensure approximate marginal stationarity within the cluster.
To identify the largest suitable value of $D_0$ that still maintains this stationarity, we apply a homogeneity test that combines the Hosking and Wallis (\citeyear{hosking1993some}) procedure with a modified Anderson–Darling test (Scholz and Stephens, \citeyear{scholz1987k}),  following the approach of \cite{castro2020local}. 

\begin{figure}[!htbp]
  \centering
  \includegraphics[width=0.6\linewidth]{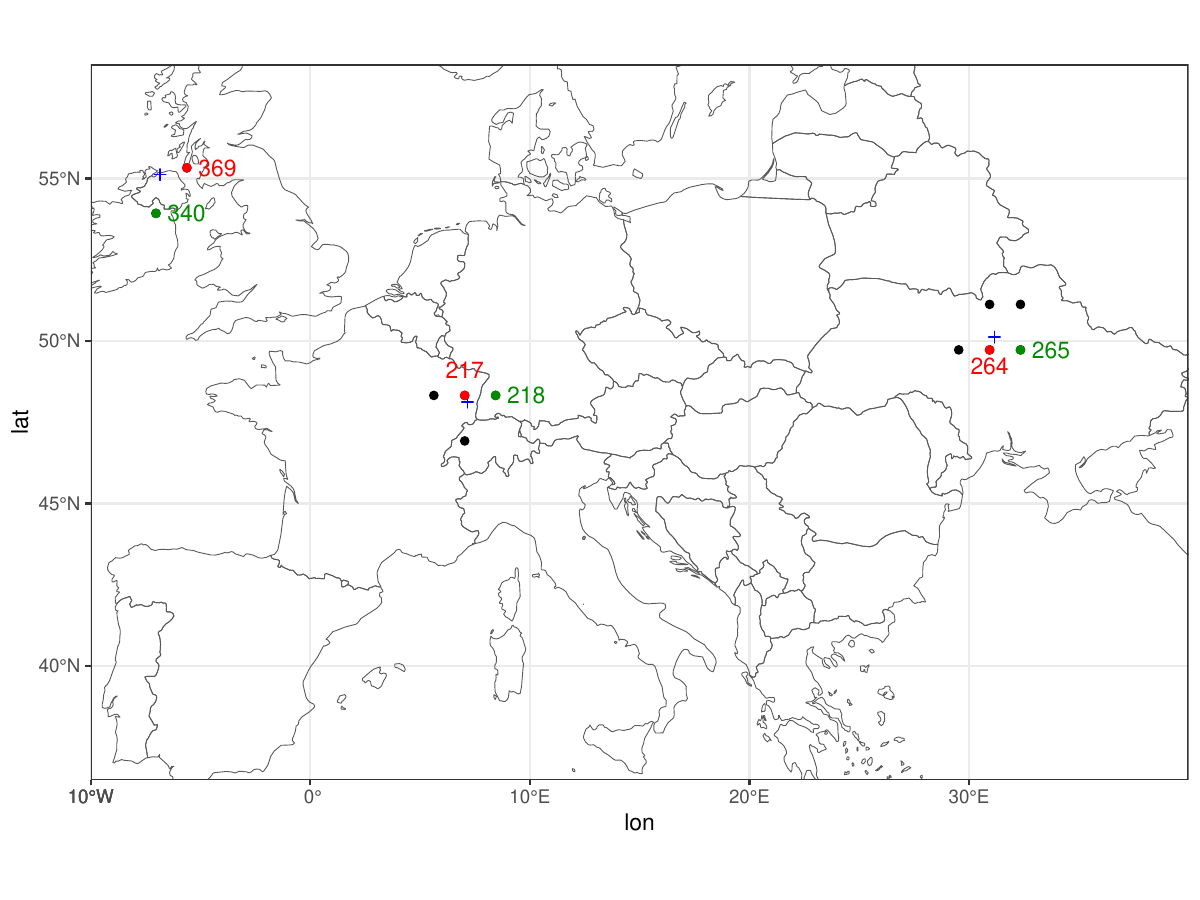}
  \caption{Study area for the first application using CNRM outputs over Europe. Blue crosses indicate three representative grid points used to define homogeneous clusters for model fitting. Within each cluster, black dots represent available observation sites, with red dots indicating the specific sites used to assess marginal fits in Figure~\ref{fig:app1_qqplot}, and green dots denoting paired sites used to evaluate dependence in Figure~\ref{fig:chi}.}
  \label{fig:map}
\end{figure}

\begin{figure}[!htbp]
  \centering
  \includegraphics[width=0.8\linewidth]{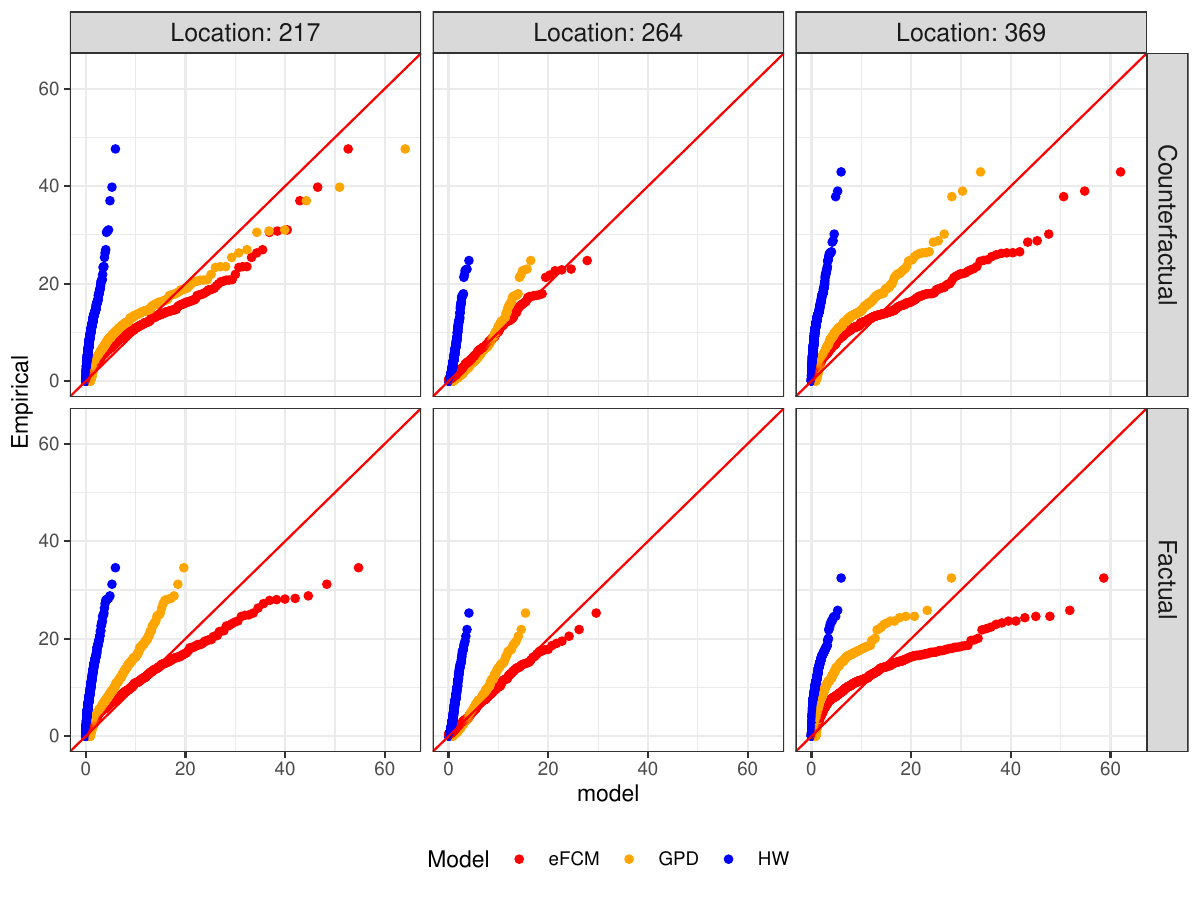}
  \caption{Marginal fit assessments for our first application (Section~\ref{sec:app1}) based on the three models, evaluated at the red-marked locations in Figure~\ref{fig:map}, using model outputs from both worlds.} 
  \label{fig:app1_qqplot}
\end{figure}

\begin{figure}[!ht]\centering
  \centering
  \includegraphics[width=0.8\linewidth]{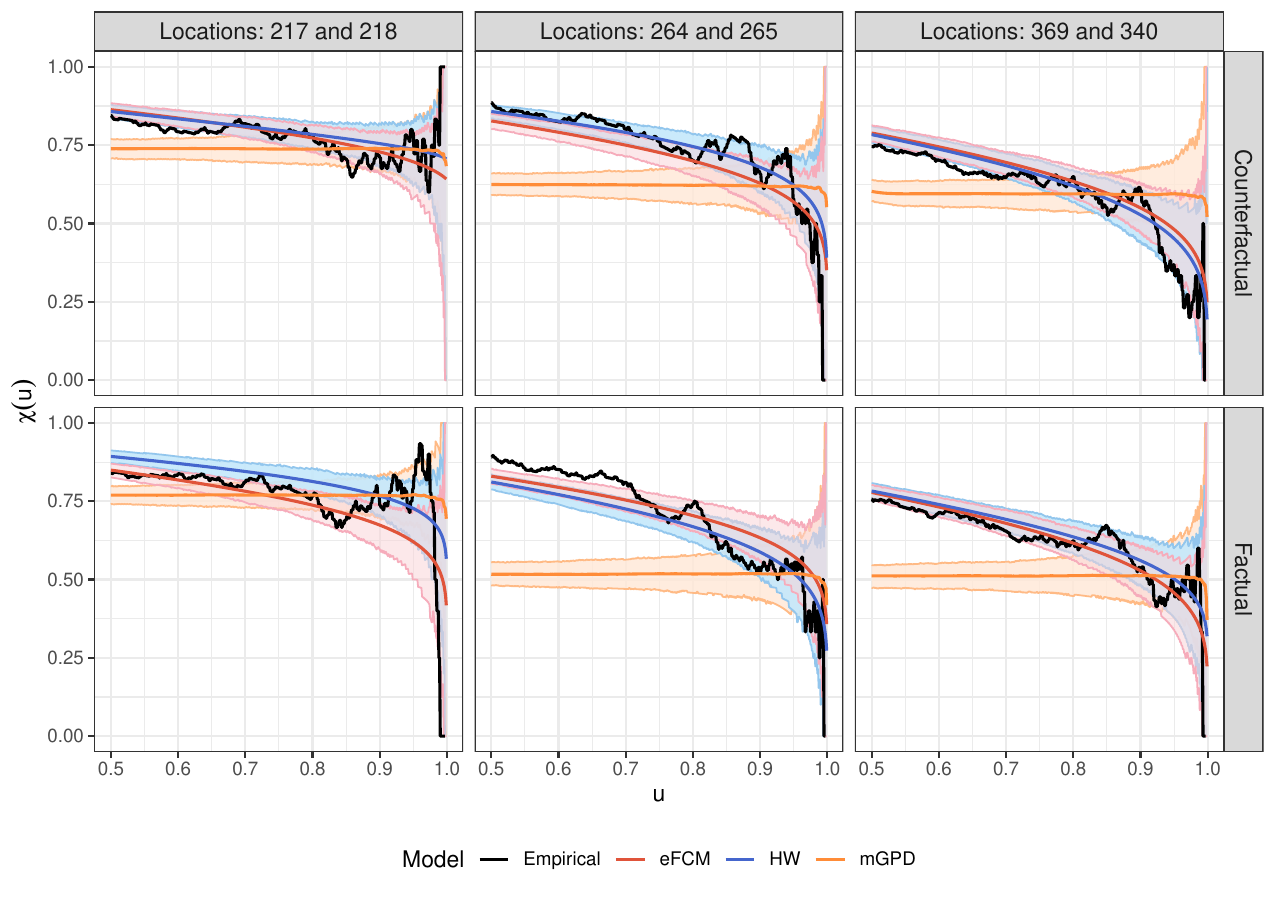}
  \caption{Conditional probabilities $\chi(u)$ as defined in Section~\ref{sec:dependence} for our first application (Section~\ref{sec:app1}) at selected locations. The black line shows the empirical estimator, while the red, blue, and orange lines correspond to the estimators from the eFCM, HW, and mGPD models, respectively. Shaded areas represent 95\% pointwise confidence intervals.}
  \label{fig:chi}
\end{figure}

Figure~\ref{fig:map} displays three example clusters identified through this procedure. 
For the three locations denoted in red (each taken from a different cluster), Figure~\ref{fig:app1_qqplot} shows marginal qqplots for the three models introduced earlier. 
The top row corresponds to the counterfactual scenario, and the bottom row corresponds to the factual scenario. 
While no model clearly dominates across all locations and worlds, the HW model consistently exhibits poorer marginal fit compared to the others.
To evaluate the performance of the dependence modeling, Figure~\ref{fig:chi} displays the bivariate conditional exceedance probability $\chi(u)$, as defined in Section~\ref{sec:dependence}, for each model. 
We focus on the three reference locations highlighted in red in Figure~\ref{fig:app1_qqplot}, pairing each with a neighbouring site from the same cluster (shown in green).
As expected, the mGPD model yields nearly constant $\chi_{\text{mGPD}}(u)$ values across levels of extremeness and pointwise estimates tend to overestimate the strength of extremal dependence. In contrast, the eFCM and HW models more closely track the empirical estimates of $\chi(u)$, indicating a better ability to capture varying tail dependence. For all models, 95\% pointwise confidence intervals are computed using nonparametric bootstrap.
Simulation studies in Section~\ref{sec:simu} show that accurate modeling of dependence is more critical than marginal fit when estimating causal quantities such as attribution ratios or probabilities of causation. 
Although both components contribute to reliable inference, dependence plays the more influential role. 
In this context, while the marginal fits in Figure~\ref{fig:app1_qqplot} show mixed performance across locations and models, the tail dependence patterns in Figure~\ref{fig:chi} suggest that the eFCM and HW models provide more realistic and robust representations of extremal dependence.
Accordingly, we place greater confidence in these models when drawing causal conclusions.

\begin{figure}[!ht]\centering
  \centering
  \includegraphics[width=\linewidth]{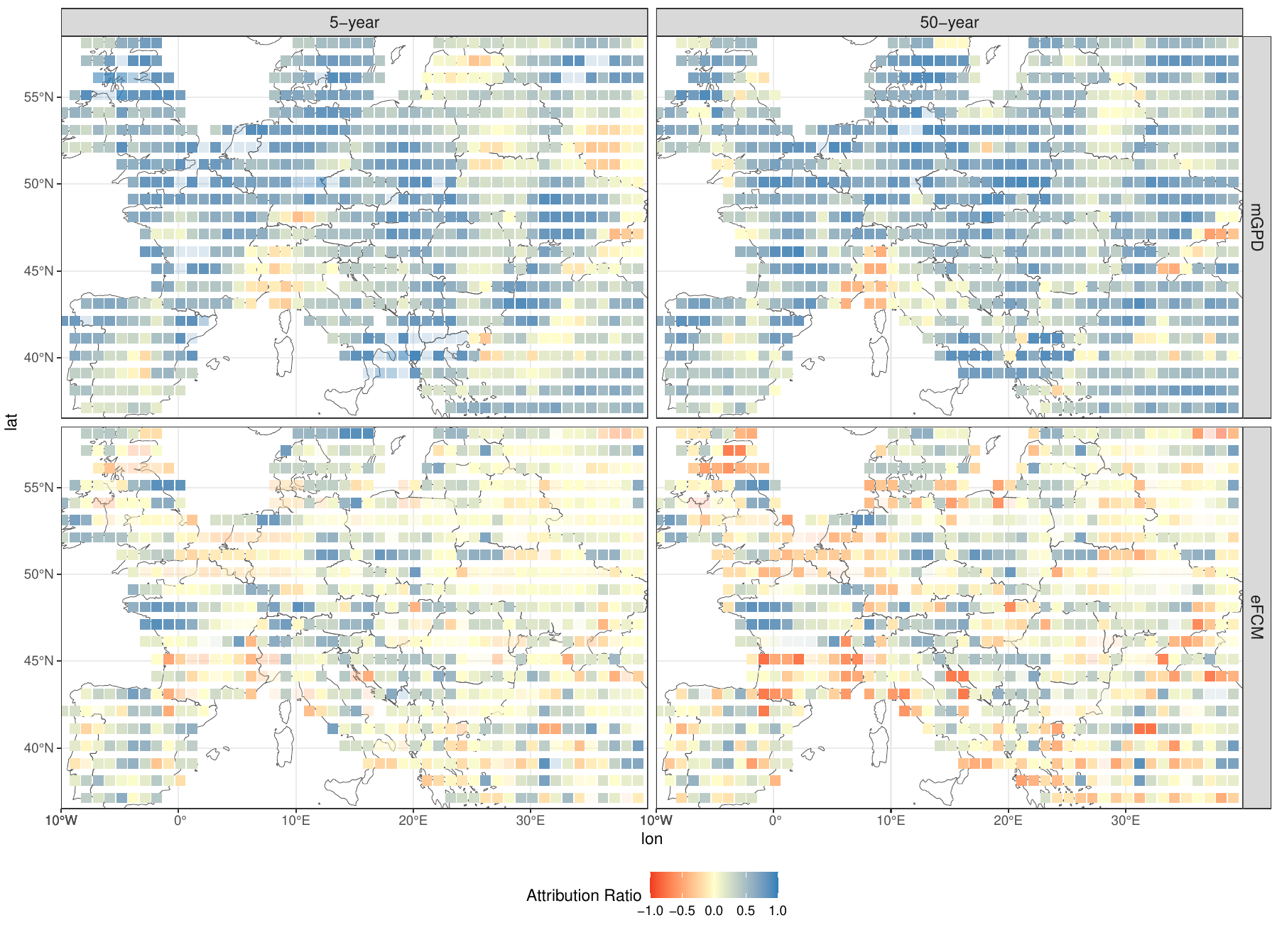}
  \includegraphics[width=\linewidth]{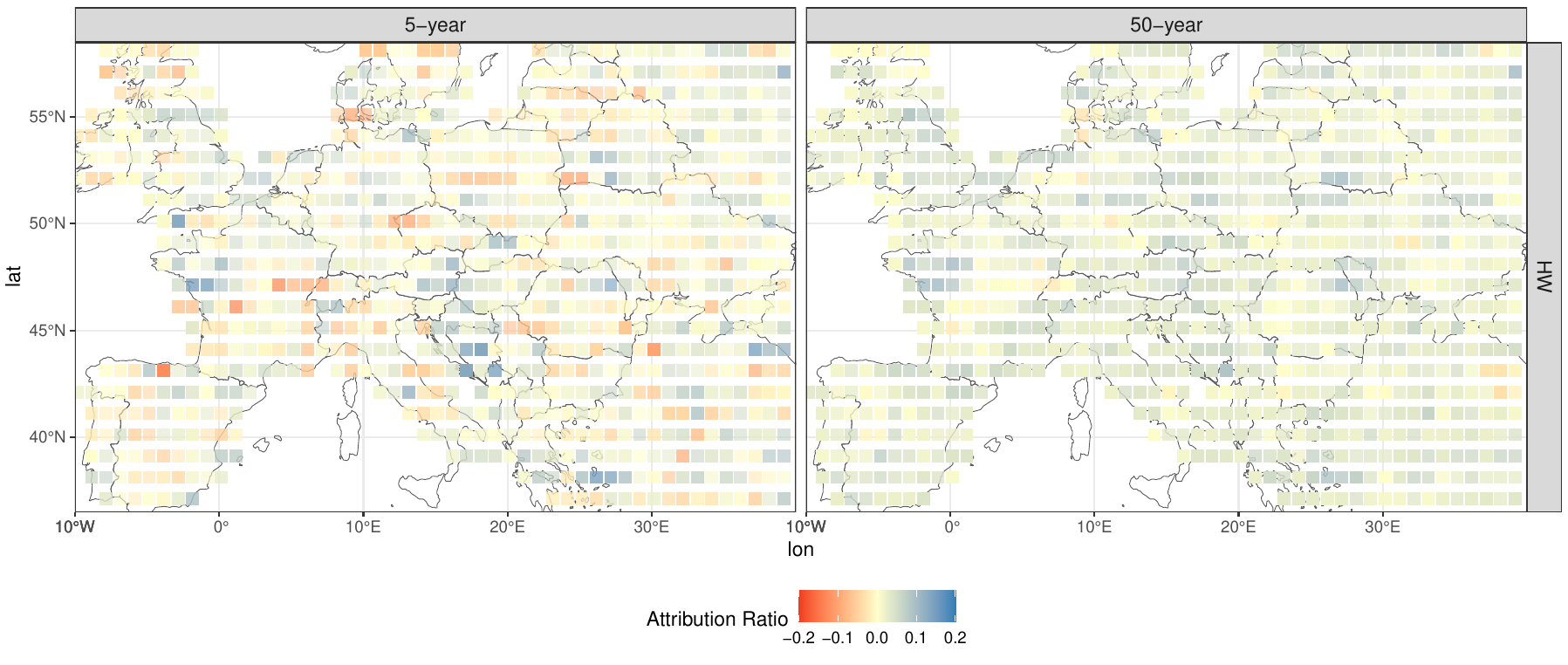}
  \caption{Spatial distribution of attribution ratios for extreme precipitation across Europe, using 5-year (left) and 50-year (right) return levels as the thresholds $v$ (see Section~\ref{sec:extremeprobs}).
The top and middle rows show results from the mGPD and eFCM models on a common $[-1,1]$ scale; the bottom row shows the HW model on its own scale due to values clustering near zero. Colors indicate AR estimates, while transparency reflects uncertainty, with more opaque colors representing greater confidence.}
  \label{fig:app1_map}
\end{figure}

Figure~\ref{fig:app1_map} displays maps of attribution ratios estimated under each of the three models, with $v$ corresponding to the 5-year (left) and 50-year (right) return levels. Colors indicate the AR estimates, while transparency reflects uncertainty: more opaque colors correspond to narrower 90\% confidence intervals and thus higher certainty, whereas greater transparency indicates increased uncertainty.
For a clearer view of the spatial patterns in AR point estimates, excluding uncertainty, see Figure~\ref{fig:appendix_pointesteurope} in Section~\ref{sec:appendix_pointesteurope} of the Appendix.
To facilitate comparison, the eFCM and mGPD results are plotted using a common color scale spanning $[-1,1]$, while the HW model uses a narrower scale of $[-0.2,0.2]$ to reflect the concentration of its estimates near zero. 
The mGPD model, which imposes a constant tail dependence structure, produces smooth and interpretable spatial patterns but, in light of the results in Figure~\ref{fig:chi}, may fail to capture localized variations in extremes. 
It also tends to yield wider confidence intervals in some regions, as evidenced by the lower opacity in its maps.
In contrast, the eFCM and HW models more effectively capture heterogeneity by incorporating flexible dependence structures through extended copula formulations.
The eFCM, in particular, consistently produces narrower confidence intervals, indicating greater statistical efficiency and stability.
With respect to spatial attribution patterns, the HW and eFCM models show broadly similar structures. 
Both indicate a mix of weakly negative to neutral ARs in Spain and Portugal, with France showing more negative values in the south and near-zero signals in the north. 
Southern Scandinavia (including Denmark and southern Sweden) shows similarly subdued and consistent patterns across both models.
Greater divergence appears in regions such as Italy, Germany, Southeast Europe, the UK, and Ireland, where the eFCM model reveals stronger negative ARs, in contrast to the mostly neutral or slightly positive estimates from the HW model.
Overall, the mGPD model suggests a decrease in risk attributable to human activity, whereas the eFCM indicates an increase, often displaying a spatial pattern similar to that of the HW model. 
Notably, regions where the mGPD detects an increase in risk, such as northern Italy and Switzerland in the 50-year return level map, are also identified by the eFCM, but are less consistently captured by the HW model.

\subsection{Precipitation in the contiguous U.S. using ACCESS-CM2 outputs} \label{sec:app2}

Our second application focuses on precipitation extremes in the contiguous U.S.
Here, we use simulations from the ACCESS-CM2 climate model under two scenarios: \emph{hist-nat}, representing a counterfactual world with only natural forcings (solar radiation and volcanic aerosols), and \emph{historical}, which includes both anthropogenic and natural forcings. 
The dataset comprises daily precipitation measurements at 251 observation location spanning 12,418 days from January 1, 1981, to December 31, 2014. 
Consistent with our first application, we compute the 7-day maxima at each location. 
However, unlike the first application, we do not assume temporal stationarity, as we now consider data spanning the entire year. 
This introduces a seasonal effect that must be accounted for. 
To address this, we use a block bootstrap approach when estimating uncertainties.

\begin{figure}[!htbp]
  \centering
  \includegraphics[width=0.8\linewidth]{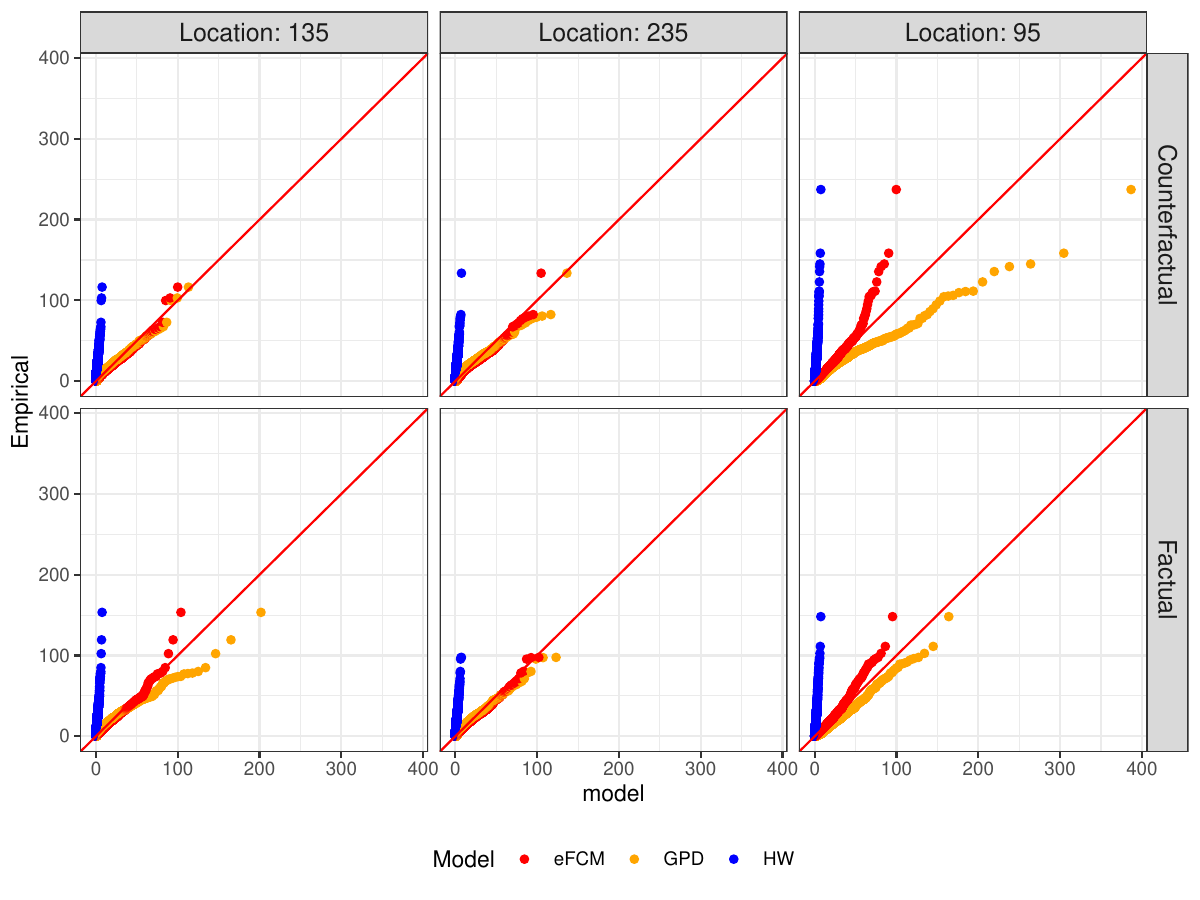}
  \caption{Marginal fit assessments for our second application (Section~\ref{sec:app2}) based on the three models, evaluated at the red-marked locations  in the bottom left panel of Figure~\ref{fig:app2_map}, using model outputs from both worlds.} 
  \label{fig:app2_qqplot}
\end{figure}

\begin{figure}[!htbp]
  \centering
  \includegraphics[width=0.8\linewidth]{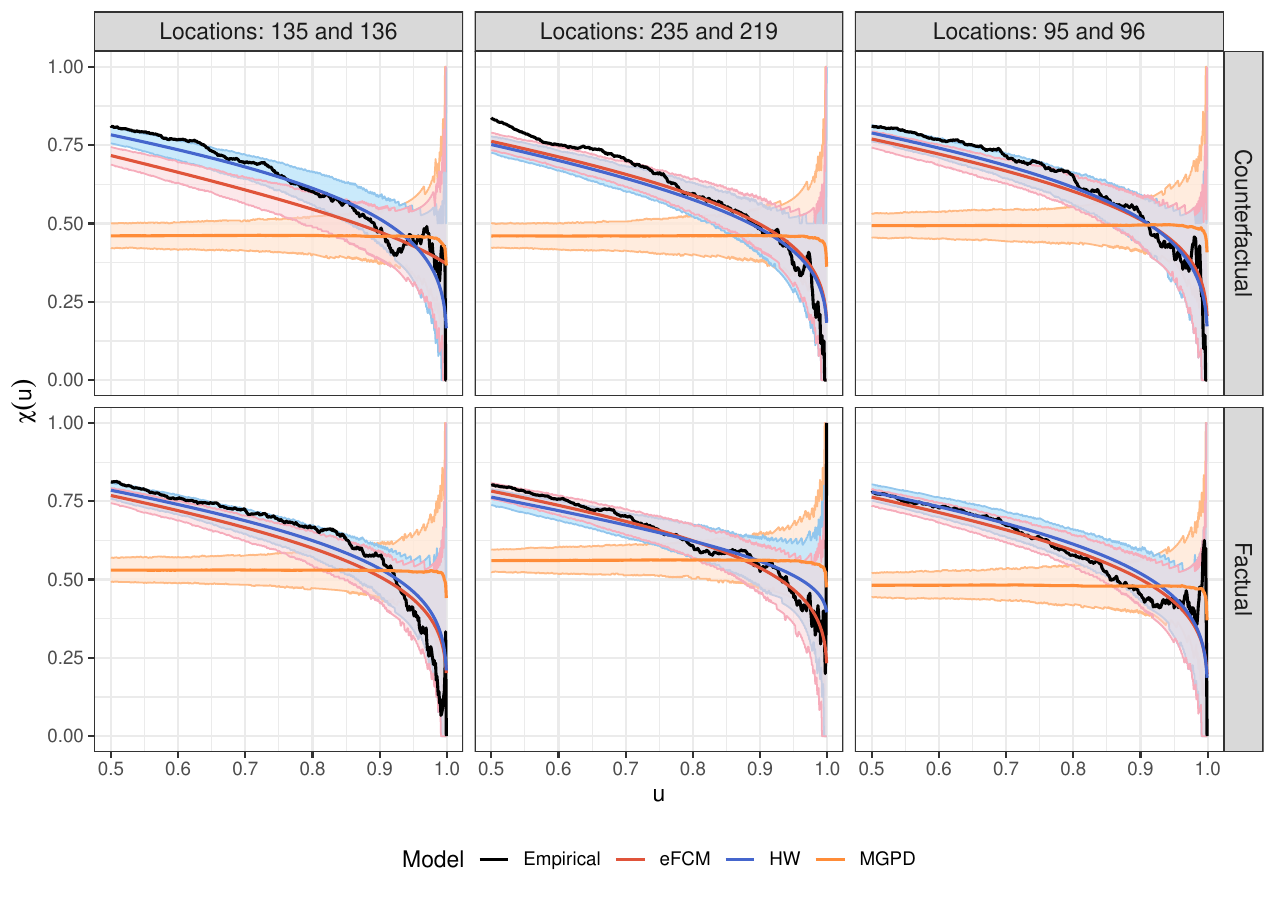}
  \caption{Conditional probabilities $\chi(u)$ as defined in Section~\ref{sec:dependence} for our second application (Section~\ref{sec:app2}) at selected locations. The black line shows the empirical estimator, while the red, blue, and orange lines correspond to the estimators from the eFCM, HW, and mGPD models, respectively. Shaded areas represent 95\% pointwise confidence intervals.} 
  \label{fig:app2_chi}
\end{figure}

\begin{figure}[!ht]\centering
  \centering
  \includegraphics[width=\linewidth]{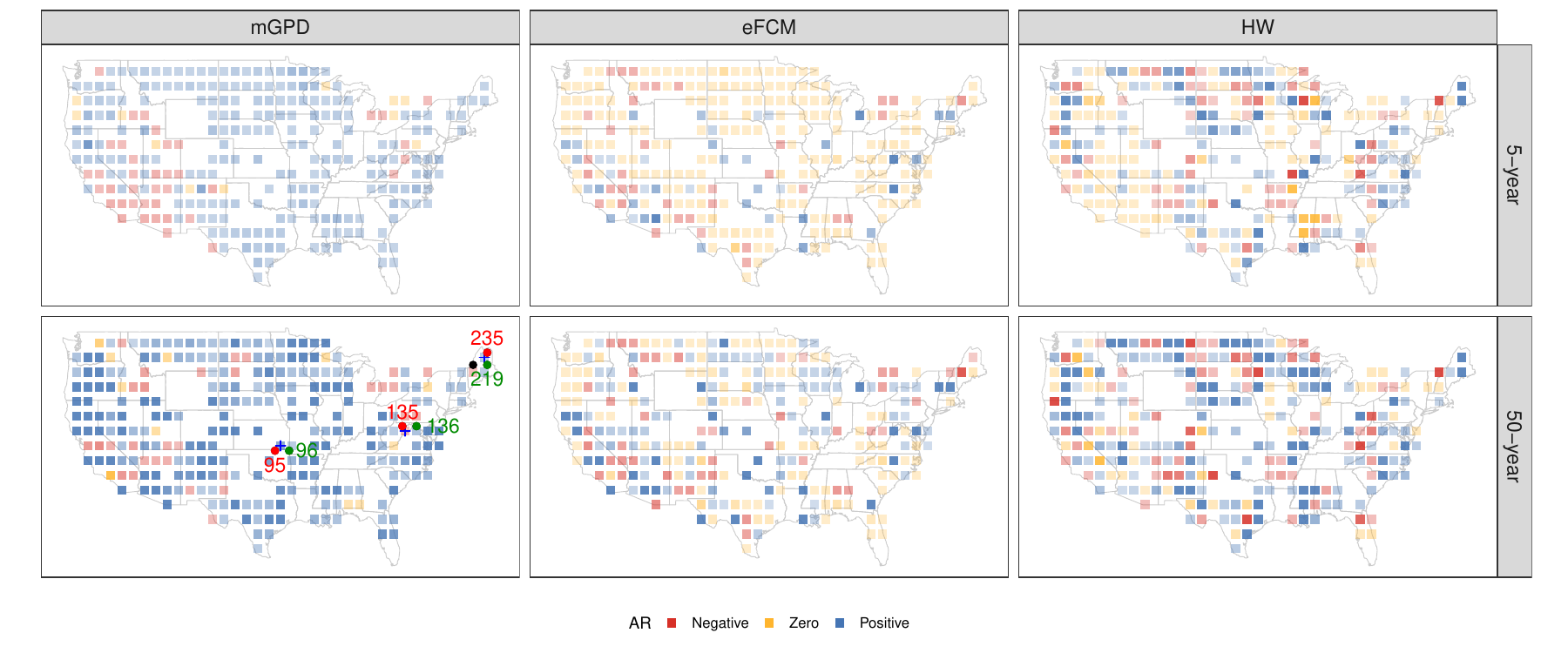}
  \caption{Spatial distribution of Attribution Ratios (AR) for extreme precipitation across the U.S, estimated using eFCM, HW, and mGPD models. The top panel shows results for the 5-year return level, and the bottom panel shows the corresponding 50-year return level estimates. Each column represents one model. AR values are categorized as negative, zero, and positive. Transparency reflects uncertainty based on the width of the 90\% confidence interval, with lighter points corresponding to higher uncertainty.}
  \label{fig:app2_map}
\end{figure}

We apply the same clustering procedure described in Section~\ref{sec:app1} to identify spatially homogeneous regions, using a grid of 261 centroids spaced at 0.5-degree intervals in both latitude and longitude.
As in the first application, we evaluate marginal fits (Figure~\ref{fig:app2_qqplot}) and tail dependence behavior (Figure~\ref{fig:app2_chi}) at the locations highlighted by red and green dots in the bottom left panel of Figure~\ref{fig:app2_map}.
For all models, 90\% confidence intervals are computed using block bootstrap with three-month blocks to account for seasonal effects.
The eFCM and mGPD models provide a reasonable fit to the marginal distributions at selected locations, while the HW model significantly underestimates the observed data. 
The performance in capturing tail dependence is consistent with the first application: both the eFCM and HW models offer better fits than the mGPD model.

To facilitate the interpretation of the causal metrics, attribution ratios displayed in Figure~\ref{fig:app2_map} are classified into three categories: \emph{negative} if they are below $-0.05$, \emph{zero} if they are less than $0.05$ in size, and \emph{positive} if they are above $0.05$. 
Confidence interval (CI) lengths are also grouped into bins of width 0.2 from 0 to 2, with a final category for lengths exceeding 2.
In the visualizations, darker colors represent narrower CIs, indicating greater certainty. 
Recall that a negative AR implies that the probability of extreme precipitation is higher under the historical (factual) scenario than under the counterfactual, suggesting an increased risk due to human influence.
The figure highlights clear differences in attribution patterns across models. 
The mGPD tends to produce strongly polarized ARs, either highly positive or highly negative, and often suggests widespread positive ARs (blue). 
This pattern, indicating a reduced probability of extremes under anthropogenic influence, may stem from model bias or oversmoothing, likely due to its assumption of constant tail dependence. 
This interpretation is supported by the model's tendency to overestimate extremal dependence, as shown in Figure~\ref{fig:app2_chi} (values of $u$ close to 1).
In contrast, both the eFCM and HW models display spatially heterogeneous AR patterns, with mixed signals across regions and consistent visual encoding of uncertainty. 
However, key differences emerge: the HW model produces stronger and more polarized attribution signals, particularly at the 50-year return level, with many regions showing confidently positive or negative ARs. 
In contrast, the eFCM model yields more conservative estimates, with a greater prevalence of near-zero and uncertain values. 
Regional discrepancies are also notable; HW shows pronounced positive ARs in the Northeast and Midwest, which are less evident under the eFCM. 
Despite these differences, some broad regional trends are shared between models, suggesting a degree of robustness in certain attribution signals while also highlighting sensitivity to model choice and event rarity.
Taken together with the results on marginal and dependence fit, these findings suggest that the eFCM model provides the most robust evidence of anthropogenic influence on extreme precipitation, especially at higher return levels. 
This is consistent with the conclusions of~\cite{nanditha2024strong} (see their Supplementary Figure S6), further supporting the reliability of the eFCM model in capturing human-driven changes in precipitation extremes.

\section{Discussion and conclusion\label{sec:conclusion}}
This study highlights the critical role that tail assumptions play in shaping causal attribution metrics in Extreme Event Attribution. 
We compared three statistical models, namely, the multivariate generalized Pareto distribution (mGPD), the exponential factor copula model (eFCM) of~\cite{castro2020local}, and the Huser-Wadsworth (HW) model~\citep{huser2019modeling}, through both simulation experiments and two real-world applications. 
 These models were chosen as representative of extreme value approaches, selected for their practical relevance and contrasting assumptions: the mGPD provides a classical threshold-based approach with relatively simple dependence modeling; the eFCM offers enhanced flexibility to capture a wide range of tail dependencies through copula-based structures; and the HW model is a flexible spatial model that bridges asymptotic dependence and asymptotic independence, allowing for the extremal dependence class to be data-driven.
 
 Our results consistently demonstrate that the choice of tail model has a substantial impact on estimated marginal distributions, dependence structure, ARs, and ultimately, on the conclusions drawn about anthropogenic influence.
Using data simulated from the HW model with known marginal and dependence structures, we show that model misspecification, particularly in the tail dependence structure, can lead to significant bias in exceedance probabilities. 
In particular, the first simulation assesses how much accurately specifying the dependence structure matters, relative to correctly modeling the marginals, for reliable inference in the tail.
In the second simulation, we vary the strength of extremal dependence and assess how well mGPD and eFCM recover the true tail probabilities. 
The eFCM consistently exhibits lower bias and RMSE across all dependence regimes, suggesting its superior flexibility and robustness in capturing tail dependence.

\begin{figure}[!htbp]
  \centering
\includegraphics[width=0.4\linewidth]{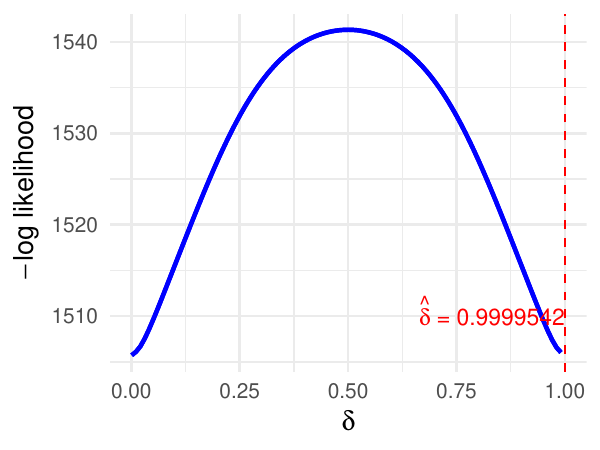}
  \caption{Profile of the marginal negative log-likelihood for $\delta$ in the marginal HW model, using data from location 217 in Application 1 (Section~\ref{sec:app1}). The curve reaches its minimum near the upper boundary of the parameter space, suggesting potential weak identifiability of $\delta$.}
  \label{fig:hw_marginal}
\end{figure}

Empirical results further reveal distinct model behaviors. 
The mGPD offers stable but potentially overconfident estimates, due to its rigid tail assumptions. 
The eFCM yields more nuanced spatial patterns with narrower confidence intervals, reflecting its adaptability to varying dependence structures. 
The HW model captures spatial variation in extremal dependence effectively, but estimation, particularly for parameters like $\delta$, can be challenging in practice. 
In both simulation and application, we observed signs of weak identifiability. 
To illustrate this, Figure~\ref{fig:hw_marginal} shows the profile of the marginal negative log-likelihood for $\delta$ at location 217 (Application 1, Section~\ref{sec:app1}; see also top left panel of Figure~\ref{fig:app1_qqplot}), with a flat peak and a minimum near the upper boundary ($\delta\approx 0.99995$), suggesting limited information in the data for precise estimation. 
While this may reflect true parameter values, estimation could be improved through penalised likelihood, reparameterisation, or Bayesian methods with informative priors.

Across both applications, we find consistent signals of anthropogenic influence when tail dependence is flexibly modeled. 
These findings highlight that tail assumptions are not merely technical choices; they play a central role in shaping attribution outcomes. 
Without careful attention to tail behavior, particularly in the dependence structure, attribution results may be misleading or overly confident.

\section*{Acknowledgements}
We thank Anna Kiriliouk for providing the processed dataset based on the CNRM output and the code to fit the mGPD model, which is now available in our \href{https://github.com/MengranLi-git/tail-eea}{GitHub repository} with her permission. The code for fitting the HW model was adapted from a version available at \url{https://www.lancaster.ac.uk/~wadswojl/SpatialADAI}.
Mengran Li gratefully acknowledges support from the China Scholarship Council (CSC).

\appendix
\section*{Appendix}
\renewcommand{\thefigure}{A\arabic{figure}}
\setcounter{figure}{0}  
\section{Expressions for bivariate tail dependence coefficients}\label{sec:appendix_chi}
Here, we provide derivations for the conditional probabilities $\chi(u)$ for the eFCM and the HW models introduced in Section~\ref{sec:methods}.

\subsection{Derivation of \(\chi(u)\) for the eFCM}\label{sec:appendix_chi_efcm}
We consider the exponential factor copula model (eFCM), where the dependence structure arises from a Gaussian copula with margins given by:
\[
W_j = Z_j + \lambda_j^{-1} E, \quad j = 1, 2,
\]
where \(Z = (Z_1, Z_2)^\top \sim \mathcal{N}(0, \Sigma)\), with \(\Sigma\) having unit variances and correlation \(\rho\), and \(E \sim \text{Exp}(1)\), independent of \(Z\).
Let \(F_j(w; \lambda_j)\) denote the marginal distribution of \(W_j\), and define
\[
z(u) = F_j^{-1}(u; \lambda_j),
\]
so that \(W_j \sim F_j\) and \(F_j(z(u); \lambda_j) = u\).
It can be shown that $\chi(u)$ can be expressed as
\[
\chi(u) = \frac{1 - 2u + C_2(u,u)}{1 - u},
\]
where $C_2$ is the bivariate copula. Using the eFCM, we have
\[
\chi_\efcm(u) = \frac{1 - 2u + C_2^W(u,u)}{1 - u},
\]
where $C_2^W$ is the eFCM bivariate copula.
From the supplementary material of \cite{castro2020local}, \(C_2^W(u,u)\) has the form:
\[
C_2^W(u, u) = \Phi_2(z_1(u), z_2(u); \rho) - \sum_{j=1}^2 \exp\left(\frac{\lambda_j^2}{2} - \lambda_j z_j(u)\right) \Phi_2(q_j, 0; \Omega_j),
\]
where \(\Phi_2(\cdot, \cdot; \rho)\) is the bivariate standard normal CDF with correlation \(\rho\), \(q_j\) and \(\Omega_j\) are functions of \(\lambda_1, \lambda_2\) and \(\rho\), provided in page 2 of the supplementary material of \cite{castro2020local}, and \(z_j(u) = F_j^{-1}(u; \lambda_j)\).
Thus, the expression for \(\chi_\efcm(u)\) becomes:
\[
\chi_\efcm(u) = \frac{1 - 2u + \Phi_2(z_1(u), z_2(u); \rho)
- \sum\limits_{j=1}^2 \exp\left(\frac{\lambda_j^2}{2} - \lambda_j z_j(u)\right) \Phi_2\left(q_j, 0; \Omega_j\right)}{1 - u}
\]
In the symmetric case (\(\lambda_1 = \lambda_2 = \lambda\)), the formula simplifies to:
\[
\chi_\efcm(u) = \frac{1 - 2u + \Phi_2(z(u), z(u); \rho)
- 2 \exp\left(\frac{\lambda^2}{2} - \lambda z(u)\right) \Phi_2\left(q, 0; \Omega\right)}{1 - u}
\]
where:
\[
q = \lambda(1 - \rho), \quad \Omega =
\begin{pmatrix}
2(1 - \rho) & \sqrt{2(1 - \rho)} \\
\sqrt{2(1 - \rho)} & 1
\end{pmatrix}.
\]

\subsection{Derivation of \(\chi(u)\) for the HW model}\label{sec:appendix_chi_hw}

Recall that in the Huser and Wadsworth (HW) model, the process is defined as $X(s) = R^\delta W(s)^{1 - \delta}, \ \delta \in (0,1),$
where \(R \sim \text{Pareto}(1)\) and \(W(s)\) is an asymptotically independent process with standard Pareto margins. Taking logarithms gives the additive form
\[
\tilde{X}(s) := \log X(s) = \delta \tilde{R} + (1 - \delta) \tilde{W}(s),
\]
where \(\tilde{R} = \log R \sim \text{Exp}(1)\), and \(\tilde{W}(s) = \log W(s)\), with \(\tilde{W}_j \sim \text{Exp}(1)\) and copula structure determined by the dependence in \(W\).

Let \(F_X\) denote the marginal distribution of \(X(s)\). Then the bivariate copula \(C_2(u,u)\) associated with the HW model is given by
\[
C_2(u,u) = \mathbb{P}\left(X_1 \le w(u), X_2 \le w(u)\right),
\]
where \(w(u) = F_X^{-1}(u)\).

From the log-scale representation, this can be written as
\[
C_2(u,u) = \mathbb{P}\left( \tilde{X}_1 \le \log w(u), \tilde{X}_2 \le \log w(u) \right) = \int_0^{\log w(u)/\delta} F_{\tilde{W}_1, \tilde{W}_2} \left( \frac{\log w(u) - \delta r}{1 - \delta}, \frac{\log w(u) - \delta r}{1 - \delta} \right) e^{-r} \, \mathrm{d}r,
\]
where \(F_{\tilde{W}_1, \tilde{W}_2}\) is the joint distribution of \((\tilde{W}_1, \tilde{W}_2)\).

Substituting into the standard formula for the conditional tail dependence function:
\[
\chi(u) = \frac{1 - 2u + C_2(u,u)}{1 - u},
\]
we obtain:
\[
\chi_\hw(u) = \frac{1 - 2u + \displaystyle \int_0^{\log w(u)/\delta} F_{\tilde{W}_1, \tilde{W}_2} \left( \frac{\log w(u) - \delta r}{1 - \delta}, \frac{\log w(u) - \delta r}{1 - \delta} \right) e^{-r} \, \mathrm{d}r }{1 - u}
\]
where \(w(u) = F_X^{-1}(u)\) is the marginal quantile function of the HW model, \(F_{\tilde{W}_1, \tilde{W}_2}\) is the bivariate distribution function of the log-transformed generator process, and \(\delta\) controls the degree of asymptotic dependence.

\section{Point estimates of ARs for precipitation in Europe using CNRM outputs}\label{sec:appendix_pointesteurope}
\begin{figure}[!ht]
  \centering
\includegraphics[width=\linewidth]{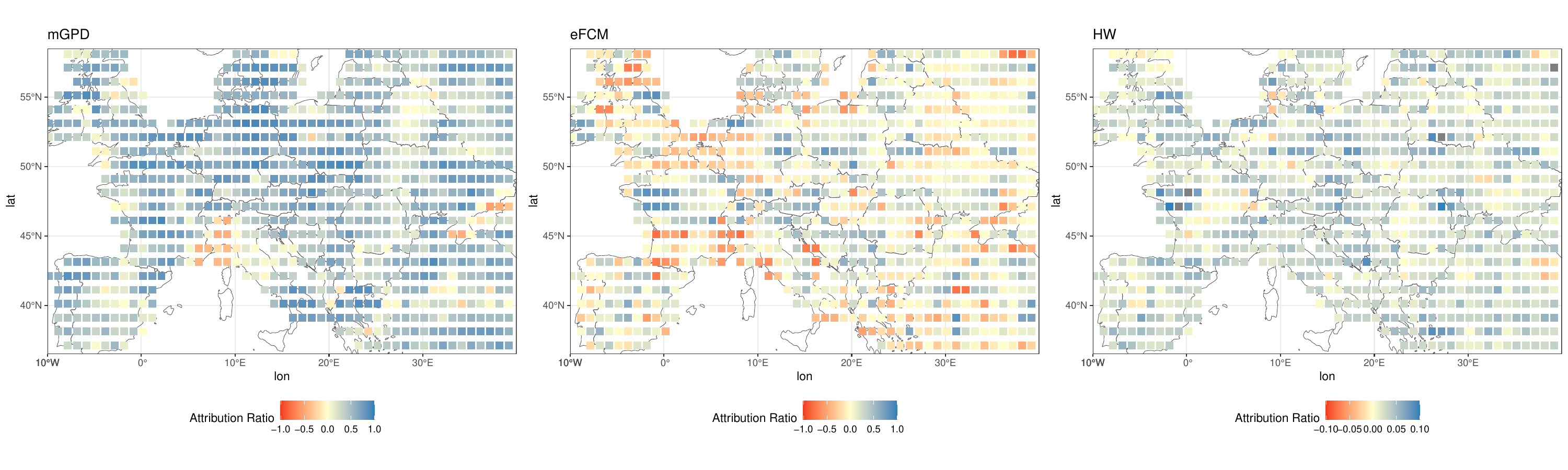}
\caption{Point estimates of attribution ratios (ARs) for the first application (Section~\ref{sec:app1}), shown without confidence intervals to highlight underlying spatial patterns across models.}
\label{fig:appendix_pointesteurope}
\end{figure}

Figure~\ref{fig:appendix_pointesteurope} presents maps of attribution ratio point estimates (excluding confidence intervals) for the first application described in Section~\ref{sec:app1}. 
By removing the visual influence of uncertainty, this simplified representation highlights the underlying spatial patterns, facilitating clearer interpretation and more direct comparison across models.

\section{Estimated PN values for both data applications}\label{sec:appendix_pn}
\begin{figure}[H]
  \centering
  \includegraphics[width=0.9\linewidth]{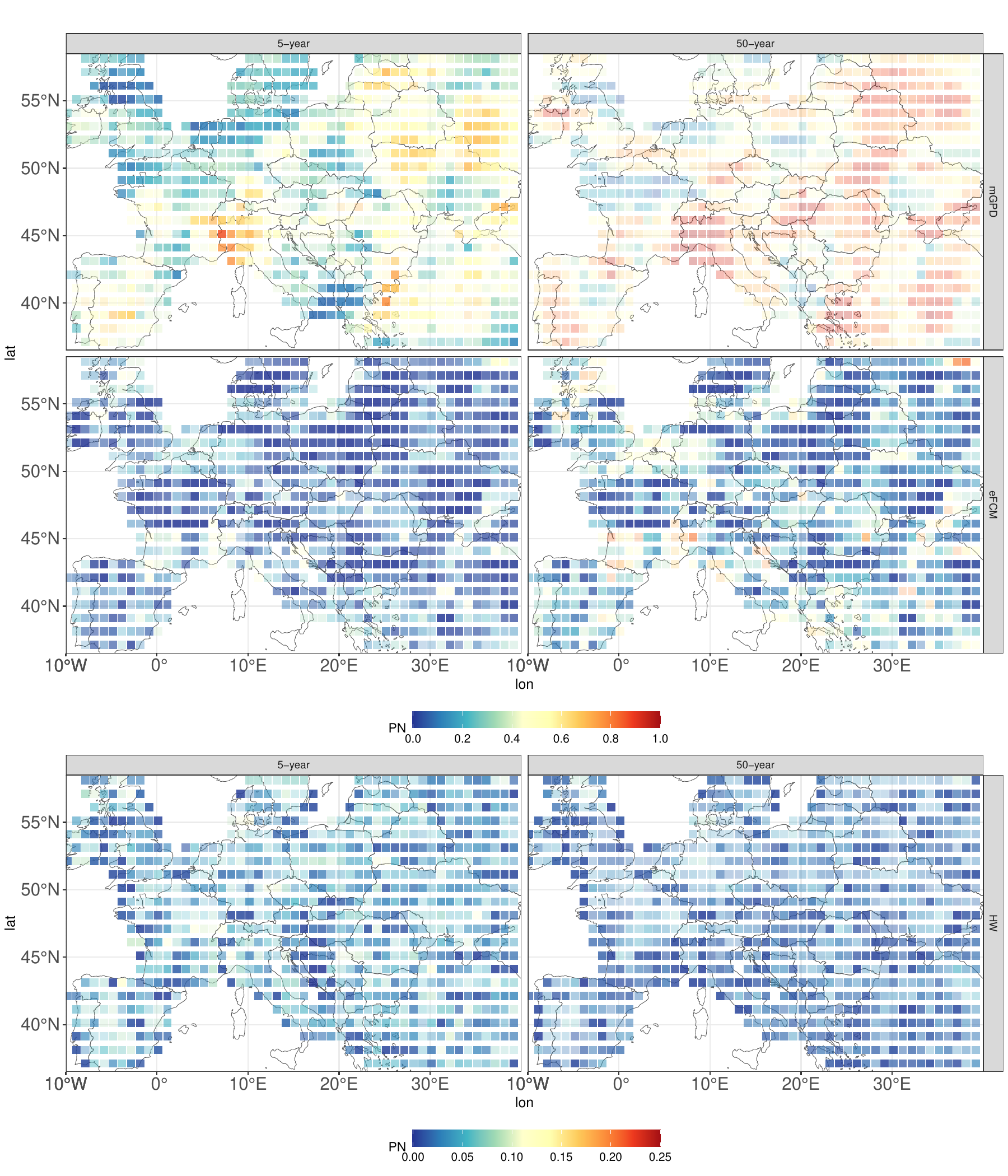}
  \caption{Spatial distribution of necessary causation probabilities for extreme precipitation across Europe, using 5-year (left) and 50-year (right) return levels as the thresholds $v$ (see Section~\ref{sec:extremeprobs}).
The top and middle rows show results from the mGPD and eFCM models on a common $[0,1]$ scale; the bottom row shows the HW model on its own scale. Solid colors are associated with estimates with narrower confidence intervals, and the color fades away as the uncertainty grows.}
  \label{fig:app1_PN}
\end{figure}

\begin{figure}[H]
  \centering
  \includegraphics[width=\linewidth]{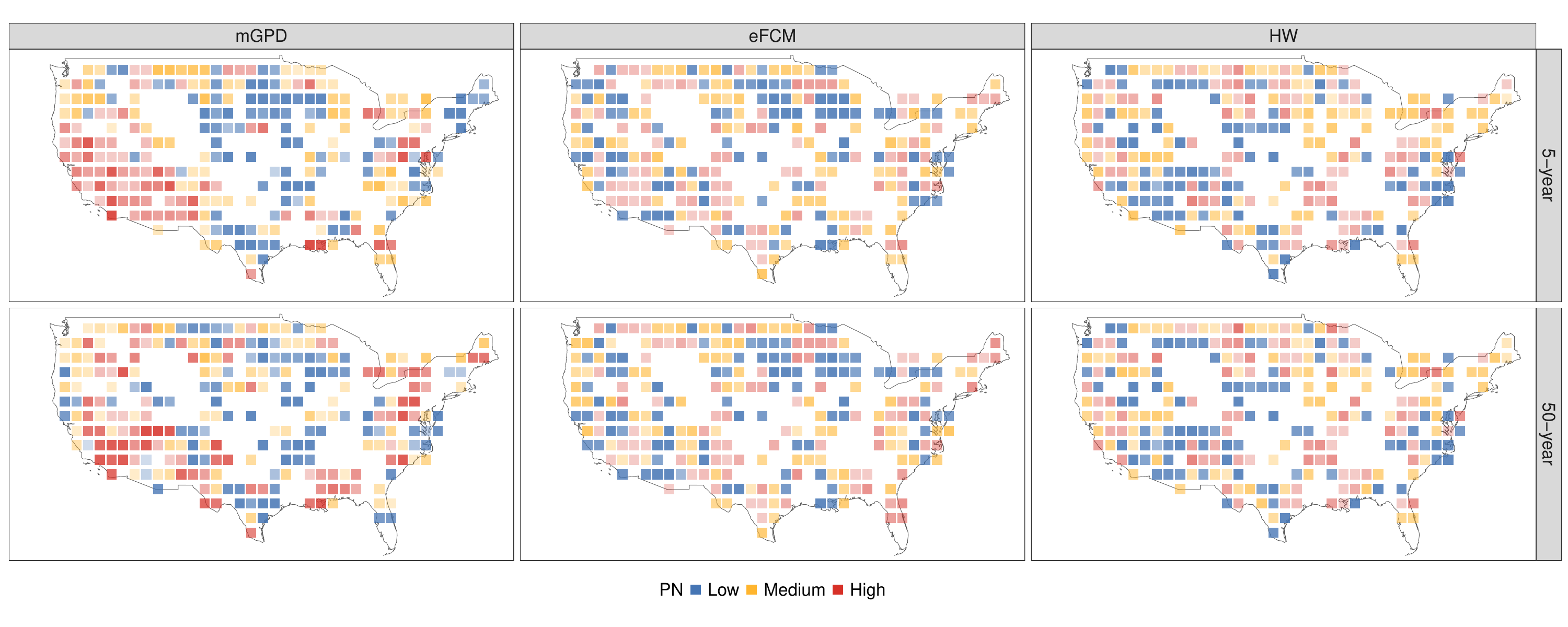}
  \caption{Spatial distribution of necessary causation probabilities for extreme precipitation across the U.S., estimated using the eFCM, HW, and mGPD models. The top panel shows results for the 5-year return level, and the bottom panel shows the corresponding 50-year return level estimates. Each column represents one model. PN values are categorized as low, medium, and high. Solid colors are associated with estimates with narrower confidence intervals, and the color fades away as the uncertainty grows.}
  \label{fig:app2_PN}
\end{figure}

As a complement to the AR maps presented in Sections~\ref{sec:app1} and~\ref{sec:app2}, we also report estimates of the probability of necessary causation (PN). 
Figure~\ref{fig:app1_PN} displays the PN estimates for the European precipitation application using CNRM outputs, while Figure~\ref{fig:app2_PN} shows the corresponding results for the contiguous U.S. application based on ACCESS-CM2 outputs. 
Both figures present results for the 5-year and 50-year return levels across the three models. 
Overall, the conclusions align closely with those drawn from the AR maps.

\newpage
\bibliographystyle{apalike}  % or another style like 'harvard', 'apalike'
\bibliography{reference.bib}

\begin{thebibliography}{}

\bibitem[Bi et~al., 2013]{bi2013access}
Bi, D., Dix, M., Marsland, S., O'Farrell, S., Rashid, H., Uotila, P., Hirst, A., Kowalczyk, E., Golebiewski, M., Sullivan, A., et~al. (2013).
\newblock The {ACCESS} coupled model: description, control climate and evaluation.
\newblock {\em Australian Meteorological and Oceanographic Journal}, 63(1):41--64.

\bibitem[Castro-Camilo and Huser, 2020]{castro2020local}
Castro-Camilo, D. and Huser, R. (2020).
\newblock Local likelihood estimation of complex tail dependence structures, applied to {US} precipitation extremes.
\newblock {\em Journal of the American Statistical Association}, 115(531):1037--1054.

\bibitem[Eyring et~al., 2016]{eyring2016overview}
Eyring, V., Bony, S., Meehl, G.~A., Senior, C.~A., Stevens, B., Stouffer, R.~J., and Taylor, K.~E. (2016).
\newblock Overview of the {C}oupled {M}odel {I}ntercomparison {P}roject {P}hase 6 ({CMIP6}) experimental design and organization.
\newblock {\em Geoscientific Model Development}, 9(5):1937--1958.

\bibitem[Gonzalez et~al., 2025]{gonzalez2025statistical}
Gonzalez, P., Naveau, P., Thao, S., and Worms, J. (2025).
\newblock A statistical method to model non-stationarity in precipitation records changes.
\newblock {\em Geophysical Research Letters}, 52(6):e2023GL107201.

\bibitem[Hannart et~al., 2016]{hannart2016}
Hannart, A., Pearl, J., Otto, F. E.~L., Naveau, P., and Ghil, M. (2016).
\newblock Causal counterfactual theory for the attribution of weather and climate-related events.
\newblock {\em Bulletin of the American Meteorological Society}, 97(1):99--110.

\bibitem[Hosking and Wallis, 1993]{hosking1993some}
Hosking, J. and Wallis, J. (1993).
\newblock Some statistics useful in regional frequency analysis.
\newblock {\em Water resources research}, 29(2):271--281.

\bibitem[Huser and Wadsworth, 2019]{huser2019modeling}
Huser, R. and Wadsworth, J.~L. (2019).
\newblock Modeling spatial processes with unknown extremal dependence class.
\newblock {\em Journal of the American Statistical Association}, 114(525):434--444.

\bibitem[Kiriliouk and Naveau, 2020]{kiriliouk2020climate}
Kiriliouk, A. and Naveau, P. (2020).
\newblock Climate extreme event attribution using multivariate peaks-over-thresholds modeling and counterfactual theory.
\newblock {\em The Annals of Applied Statistics}, 14(3):1342--1358.

\bibitem[Kiriliouk et~al., 2019]{kiriliouk2019peaks}
Kiriliouk, A., Rootz{\'e}n, H., Segers, J., and Wadsworth, J.~L. (2019).
\newblock Peaks over thresholds modeling with multivariate generalized {P}areto distributions.
\newblock {\em Technometrics}, 61(1):123--135.

\bibitem[Masson-Delmotte et~al., 2021]{masson2021climate}
Masson-Delmotte, V., Zhai, P., Pirani, A., Connors, S.~L., P{\'e}an, C., Berger, S., Caud, N., Chen, Y., Goldfarb, L., Gomis, M.~I., et~al. (2021).
\newblock Climate {C}hange 2021: The {P}hysical {S}cience {B}asis. {C}ontribution of working group {I} to the {S}ixth {A}ssessment {R}eport of the {I}ntergovernmental {P}anel on {C}limate {C}hange.
\newblock {\em IPCC: Geneva, Switzerland}.

\bibitem[McPhillips et~al., 2018]{mcphillips2018defining}
McPhillips, L.~E., Chang, H., Chester, M.~V., Depietri, Y., Friedman, E., Grimm, N.~B., Kominoski, J.~S., McPhearson, T., M{\'e}ndez-L{\'a}zaro, P., Rosi, E.~J., et~al. (2018).
\newblock Defining extreme events: A cross-disciplinary review.
\newblock {\em Earth's Future}, 6(3):441--455.

\bibitem[Nanditha et~al., 2024]{nanditha2024strong}
Nanditha, J., Villarini, G., Kim, H., and Naveau, P. (2024).
\newblock Strong linkage between observed daily precipitation extremes and anthropogenic emissions across the contiguous {U}nited {S}tates.
\newblock {\em Geophysical Research Letters}, 51(20):e2024GL109553.

\bibitem[Naveau et~al., 2020]{Naveau2020}
Naveau, P., Hannart, A., and Ribes, A. (2020).
\newblock Statistical methods for extreme event attribution in climate science.
\newblock {\em Annual Review of Statistics and Its Application}, 7(1):89--110.

\bibitem[Paciorek et~al., 2018]{paciorek2018quantifying}
Paciorek, C.~J., Stone, D.~A., and Wehner, M.~F. (2018).
\newblock Quantifying statistical uncertainty in the attribution of human influence on severe weather.
\newblock {\em Weather and climate extremes}, 20:69--80.

\bibitem[Pearl, 2009]{pearl2009}
Pearl, J. (2009).
\newblock Causal inference in statistics: An overview.
\newblock {\em Statistics Surveys}, 3:96 -- 146.

\bibitem[Rootz{\'e}n et~al., 2018]{rootzen2018multivariate}
Rootz{\'e}n, H., Segers, J., and L.~Wadsworth, J. (2018).
\newblock Multivariate peaks over thresholds models.
\newblock {\em Extremes}, 21(1):115--145.

\bibitem[Rootz{\'e}n and Tajvidi, 2006]{rootzen2006multivariate}
Rootz{\'e}n, H. and Tajvidi, N. (2006).
\newblock Multivariate generalized {P}areto distributions.
\newblock {\em Bernoulli}, 12(5):917--930.

\bibitem[Scholz and Stephens, 1987]{scholz1987k}
Scholz, F.~W. and Stephens, M.~A. (1987).
\newblock K-sample {A}nderson--{D}arling tests.
\newblock {\em Journal of the American Statistical Association}, 82(399):918--924.

\bibitem[Shadish et~al., 2002]{shadish2002}
Shadish, W., Cook, T., and Campbell, D. (2002).
\newblock {\em Experimental and Quasi-Experimental Designs for Generalized Causal Inference}.
\newblock Cengage Learning, 2 edition.

\bibitem[Van Der~Wiel et~al., 2017]{van2017rapid}
Van Der~Wiel, K., Kapnick, S.~B., Van~Oldenborgh, G.~J., Whan, K., Philip, S., Vecchi, G.~A., Singh, R.~K., Arrighi, J., and Cullen, H. (2017).
\newblock Rapid attribution of the {A}ugust 2016 flood-inducing extreme precipitation in south {L}ouisiana to climate change.
\newblock {\em Hydrology and Earth System Sciences}, 21(2):897--921.

\bibitem[van Oldenborgh et~al., 2021]{van2021pathways}
van Oldenborgh, G.~J., van~der Wiel, K., Kew, S., Philip, S., Otto, F., Vautard, R., King, A., Lott, F., Arrighi, J., Singh, R., et~al. (2021).
\newblock Pathways and pitfalls in extreme event attribution.
\newblock {\em Climatic Change}, 166(1):1--27.

\bibitem[Voldoire et~al., 2019]{voldoire2019evaluation}
Voldoire, A., Saint-Martin, D., S{\'e}n{\'e}si, S., Decharme, B., Alias, A., Chevallier, M., Colin, J., Gu{\'e}r{\'e}my, J.-F., Michou, M., Moine, M.-P., et~al. (2019).
\newblock Evaluation of {CMIP6} deck experiments with {CNRM-CM6-1}.
\newblock {\em Journal of Advances in Modeling Earth Systems}, 11(7):2177--2213.

\bibitem[{World Economic Forum}, 2022]{wef2022}
{World Economic Forum} (2022).
\newblock The {G}lobal {R}isks {R}eport 2022 (17th ed.).
\newblock \url{https://www3.weforum.org/docs/WEF_The_Global_Risks_Report_2022.pdf}.

\end{thebibliography}

\end{document}